\documentclass[useAMS,usenatbib]{mn2e}

\usepackage{graphicx}

\title[Testing the fragmentation limit in Upper Sco]{
Testing the fragmentation limit in the Upper Sco association
\thanks{Based on observations collected with the United Kingdom InfraRed
Telescope, Canada France Hawaii Telescope, and Very Large Telescope.}}
\author[N. Lodieu]{N. Lodieu$^{1,2}$\thanks{E-mail: nlodieu@iac.es},
\& N. C. Hambly$^{3}$, P. D. Dobbie$^{4}$, N. J. G. Cross$^{3}$, 
L. Christensen$^{5}$,
 \newauthor
E. L. Martin$^{6}$, and L. Valdivielso$^{7,1,2}$ \\
$^{1}$ Instituto de Astrof\'isica de Canarias (IAC), C/ V\'ia L\'actea s/n, 
E-38200 La Laguna, Tenerife, Spain \\
$^{2}$ Departamento de Astrof\'isica, Universidad de La Laguna (ULL),
E-38205 La Laguna, Tenerife, Spain \\
$^{3}$ Scottish Universities Physics Alliance (SUPA), Institute for Astronomy, School of Physics and Astronomy, \\
University of Edinburgh, Royal Observatory, Blackford Hill, Edinburgh EH9 3HJ, UK \\
$^{4}$ Australian Astronomical Observatory, PO Box 296, Epping, NSW, 1710, 
Australia \\
$^{5}$ Excellence Cluster Universe, Technische Universit\"at M\"unchen, 
Bolzmanstrasse 2, 85748 Garching bei M\"unchen, Germany \\
$^{6}$ CSIC-INTA Centro de Astrobiolog\'ia, Ctra. Ajalvir km 4, 28850, Torrej\'on de Ardoz, Madrid, Spain \\
$^{7}$ Centro de Estudios de F\'isica del Cosmos de Arag\'on (CEFCA), Plaza 
San Juan, 1, E-44001 Teruel, Spain  \\
}
\begin{document}

\date{Accepted \today. Received \today; in original form \today}

\pagerange{\pageref{firstpage}--\pageref{lastpage}} \pubyear{2005}

\maketitle

\label{firstpage}

%
%
\begin{abstract}
We present the results of a deep ($J$ $\sim$ 21 at 5$\sigma$) infrared
photometric survey of a 0.95 square degree area in the central region of 
the Upper Sco association. The photometric observations consist of a deep 
$Y+J$-band images obtained with the WFCAM camera on the UKIRT InfraRed 
Telescope (UKIRT) with partly coverage in $Z$ complemented by methane ON 
and OFF conducted with WIRCam on the Canada France Hawaii Telescope. 
We have selected five potential T--type objects belonging to the Upper Sco
association on the basis of their blue methane colours and their 
$J-CH_{\rm 4\,off}$ colours. We have also identified a sample of 7--8 
Upper Sco member candidates bridging the gap between known cluster M--types 
and our new T--type candidates. These candidates were selected based on
their positions in various colour--magnitude diagrams and they follow the 
sequence of known Upper Sco members identified in the UKIRT Infrared Deep 
Sky Survey (UKIDSS) Galactic Clusters Survey (GCS). We present additional 
membership constraints using proper motion estimates from the multiple
epochs available to us. We also present optical and near--infrared spectra 
obtained with the X--Shooter spectrograph on the Very Large Telescope for
five L--type candidates covering the 0.6 to 2.5 micron wavelength 
range, none of them being confirmed as a young brown dwarf. We discuss the
lack of detection of new candidate members as well as the possible turn down
in the USco mass function as we are approaching the fragmentation limit.
\end{abstract}

\begin{keywords}
Stars: low-mass stars and brown dwarfs --- techniques: photometric --- 
techniques: spectroscopic --- Infrared: Stars  --- surveys ---
stars: luminosity function, mass function
\end{keywords}

%
%
\section{Introduction}
\label{deepUsco:intro}

Knowledge of the shape of the present--day mass function and the masses 
of the coolest substellar objects in young clusters and star forming regions 
is of prime importance to put constraints on the formation mechanisms 
responsible for the existence of brown dwarfs and planetary mass objects.
According to the standard theory of star formation, the fragmentation limit,
i.e.~the mass at which one object is unable to contract further because it 
can not radiate its heat to collapse further, is of the order of a few 
Jupiter masses \citep[5--10 M$_{\rm Jup}$;][]{low76,rees76}. However, this 
limit might be lower, possibly of the order of $\sim$1 M$_{\rm Jup}$ in 
presence of magnetic fields \citep{boss01}, but more realistic 3D 
calculations of the collapse of molecular clouds are required to thoroughly 
address this issue from the theoretical point of view. 

Observationally, several regions have been probed to faint magnitudes at both 
optical and near--infrared wavelengths to look for a possible cut--off at the 
low mass end of the mass function. \citet{bihain10a} reported a possible 
cut--off below $\sim$6 M$_{\rm Jup}$ in the mass function of the 
1--8 Myr--aged $\sigma$ Orionis cluster from a deep ($J$ $\sim$ 21.5 mag) 
optical/infrared survey over 0.23 square degree. These authors added one 
further very young objects pre-emptively classified as a T--type candidate
to the previously known S\,Ori\,70. The latter object has a T6 spectral type 
and displays unusual infrared colours pointing towards a younger age 
\citep{zapatero02b,zapatero08a} but its membership of $\sigma$ Orionis 
remains a matter of debate \citep{burgasser04b,scholz08a,luhman08c,penya11a}.
\citet{alves10} presented spectroscopy for a subsample of low mass brown dwarf 
candidates identified in a survey extending over the full projected area of 
the $\rho$ Ophiuchus cloud and reaching to $J$ $\sim$ 20.5 mag. These authors 
derive luminosities and effective temperatures for their candidate members 
but do not present a mass function. Independently, \citet{geers11} has 
confirmed through infrared
spectroscopy several candidates as substellar members including one at 
the deuterium burning limit. \citet{marsh10} published the discovery of 
a T2 dwarf in the same region, finding not confirmed by 
\citet{alves10}. \citet{scholz09a} reported, on the basis of a deep optical 
and infrared survey of 0.25 square degrees in NGC\,1333 complemented by 
spectroscopy, a possible dearth of planetary mass objects corresponding to 
a potential cut--off in the cluster mass function in the 20--12 M$_{\rm Jup}$ 
mass range. \citet{burgess09} identified a mid-T--type candidate from a 
methane survey of $\sim$0.11 square degrees in IC\,348, suggesting that
an extrapolation of the field mass function \citep{chabrier03} may
hold in the substellar and planetary mass regimes. \citet{casewell07}
extracted several L/T--type candidates in 2.5 square degree in the
Pleiades cluster from deep optical and infrared photometry complemented
by proper motion using the 5 year baseline between both surveys.
Evolutionary models \citep{baraffe02} predict masses as low as 
10 M$_{\rm Jup}$ for these candidates if indeed members of the Pleiades.
The mass function appears to keep rising at such low masses with a slope index
which agrees, within the uncertainties, with the values inferred from earlier 
studies \citep[e.g.][]{dobbie02a,moraux03,lodieu07c}.

The Upper Sco association (hereafter USco) is part of the Scorpius
Centaurus OB association: it is located at 145 pc from the Sun
\citep{deBruijne97} and its age is estimated to 5$\pm$2 Myr from isochrone
fitting and dynamical studies \citep{preibisch02}. The association was
targeted in X~rays \citep{walter94,kunkel99,preibisch98}, astrometrically
ith Hipparcos \citep{deBruijne97,deZeeuw99}, and more recently at optical
\citep{preibisch01,preibisch02,ardila00,martin04,slesnick06}
and near--infrared \citep{lodieu06,lodieu07a} wavelengths.
Several tens of brown dwarfs have now been confirmed spectroscopically 
as members of the association
\citep{martin04,slesnick06,lodieu06,slesnick08,lodieu08a,martin10a,lodieu11a}
and the mass function of this population determined well into the substellar 
regime \citep{slesnick08,lodieu11a}.

In this paper we report the outcome of a deep photometric survey of
approximately 0.95 square degrees in the central region of USco to
(1) find young T--types, (2) test the theory of the fragmentation limit 
in a young nearby star forming region, and (3) bridge the gap between
previous surveys and the new T--type candidates. 
In Section~\ref{deepUsco:phot_obs} we describe methane ON and OFF imaging 
survey obtained with the Canada France Hawaii Telescope (CFHT), the 
near-infrared $ZYJ$ imaging survey carried out with the United Kingdom (UK) 
InfraRed Telescope (UKIRT), the optical $z$-band imaging obtained
with the Wide-Field Camera (WFC) on the Isaac Newton Telescope (INT) and
the optical to near-infrared spectroscopy obtained with the X-Shooter
instrument on the Very Large Telescope.
In Section~\ref{deepUsco:select_Tdwarfs} we outline the photometric selection
of T--type candidates using the methane survey complemented by the deep
optical $z$~band. 
In Section~\ref{deepUsco:select_Ldwarfs} we attempt to bridge the gap between
our previous studies of USco and the search for T--types presented in this
paper using various colour--magnitude diagrams drawn from all datasets 
presented here. We also make use of the multiple epochs to estimate the 
proper motion of some of the new candidates.
In Section~\ref{deepUsco:XSHOOTERspec} we present the X-Shooter spectra of
several photometric candidates covering the optical to near--infrared 
(0.6--2.2 microns) wavelength range.
We summarise our investigation in Section~\ref{deepUsco:summary} and
propose future avenues for exploration.

%
%
\section{Imaging observations}
\label{deepUsco:phot_obs}

In this section, we describe methane ON and OFF, deep $ZYJ$, and optical $z$
photometric observations aiming at identifying potential substellar members 
in USco.

\subsection{CFHT WIRCam methane survey}
\label{deepUsco:phot_obs_CFHT}
\subsubsection{WIRCam Observations}

The Wide--field InfraRed Camera \citep[WIRCam;][]{puget04} is a near--infrared 
(0.9--2.4 microns) camera installed on the prime focus of the 3.6m CFHT 
on Mauna Kea, Hawaii. WIRCam is equipped with four 2048$\times$2048 HAWAII 2RG 
detectors. The pixel scale is 0.3 arcsec, thus yielding an almost contiguous 
field-of-view of 20.5 arcmin aside. We performed observations in February,
April, June, and July 2008 with the methane OFF and ON filters which are centered 
at 1.58 and 1.69 microns, respectively, and covered nine pointings (i.e.\ one 
square degree) which are co-incident with our deep $YJ$ imaging (the 
overlapping region is exactly 0.95 square degrees). The centre of the WIRCam 
coverage is located at 16$^{h}$ 08$^{m}$,$-$22$^\circ$ 45', equinox J2000.0\@. 
The total time per WIRCam field, including overheads, amounted to 72 min. 
We used a five point dither pattern with on-source single integrations of 38 
sec (see Table~\ref{tab_deepUsco:log_CFHT}). The exceptions are for the tiles 
numbers 5 and 6; our subsequent strategy was optimised to better monitor sky 
variations at infrared wavelengths. Seeing conditions were generally between 
0.6 and 0.8 arcsec at zenith in the $K$--band. Dome flats in the methane OFF 
and ON filters were taken as part of our program.

%
%
%
\begin{table*}
 \centering
 \caption[]{Log of the CFHT WIRCam photometric observations.}
 \begin{tabular}{c c c c c c}
 \hline
 \hline
Tile     &   R.A.      &   dec     &  CH$_{\rm 4on}$        & CH$_{\rm 4off}$ &  Date \\
 \hline
USco\_M1  &  16:06:40  & $-$23:05  &  5$\times$14$^{\rm m}$24s & 5$\times$14$^{\rm m}$24 & 18/02, 19$+$21/06, 12$+$15/07/2008 \\
USco\_M2  &  16:06:40  & $-$22:45  &  5$\times$14$^{\rm m}$24s & 5$\times$14$^{\rm m}$24 & 24/06, 10$+$15/07/2008 \\
USco\_M3  &  16:06:40  & $-$22:25  &  5$\times$14$^{\rm m}$24s & 5$\times$14$^{\rm m}$24 & 15$+$16/07/2008 \\
USco\_M4  &  16:08:00  & $-$23:05  &  5$\times$14$^{\rm m}$24s & 5$\times$14$^{\rm m}$24 & 16$+$17/07/2008 \\
USco\_M5  &  16:08:00  & $-$22:45  &  1$\times$1$^{\rm h}$12s  & 1$\times$1$^{\rm h}$12  & 20$+$21/02/2008 \\
USco\_M6  &  16:08:00  & $-$22:25  &  1$\times$1$^{\rm h}$12s  & 5$\times$14$^{\rm m}$24 & 21$+$25/02, 20$+$21/04/2008 \\
USco\_M7  &  16:09:20  & $-$23:05  &  5$\times$14$^{\rm m}$24s & 5$\times$14$^{\rm m}$24 & 17$+$18/07/2008\\
USco\_M8  &  16:09:20  & $-$22:45  &  5$\times$14$^{\rm m}$24s & 5$\times$14$^{\rm m}$24 & 11$+$12/07/2008\\
USco\_M9  &  16:09:20  & $-$22:25  &  5$\times$14$^{\rm m}$24s & 5$\times$14$^{\rm m}$24 & 22$+$24/04, 18$+$24$+$24/06/2008 \\
  \hline
 \label{tab_deepUsco:log_CFHT}
 \end{tabular}
\end{table*}
\subsubsection{Data reduction}

The data processing of the WIRCam images is done in two steps. First, the
detrending (or removal of instrument imprints), done with the ``I''iwi 
pipeline, provides raw and detrended images for our CFHT/WIRCam run.
The second stage which involves sky-subtraction, astrometry 
and photometry is performed with SExtractor \citep{bertin96} at 
Terapix\footnote{http://terapix.iap.fr/} and outputs fully photometrically and 
astrometrically calibrated catalogues. 

Our observing program produced 729 data cubes, including 702 containing 
two images and 27 composed of 8 images. The 729 cubes are divided up into 387 
and 342 cubes in methane OFF and ON, respectively. However, several images 
(roughly 100 or 0.06\% of the total number of images) had to be removed at 
different stages of the processing for various reasons, e.g.~ large extinction due to cloud, 
the presence of strong electronic noise stripes or other instrument malfunctions.

The detrending steps involve several corrections to remove the cosmetics
of the WIRCam infrared detector\footnote{More details on the pre-processing
of WIRCam images at 
www.cfht.hawaii.edu/Instruments/Imaging/WIRCam/IiwiVersion1Doc.html} and 
includes flagging of the saturated pixels, non-linearity and reference pixel 
corrections, dark and flat-field subtraction, bad pixels removal and guiding window 
masking. The resulting images are then processed by Terapix, following closely the 
sequence and data flow chart outlined below:

\begin{enumerate}
\item Creation of masks to identify saturated pixels and reject
saturated sources
\item Field to field relative astrometric and photometric calibration.
The initial global astrometric calibration is based on the 2MASS point
source catalogue \citep{cutri03,skrutskie06}
\item Generate a first stack with the detrended and sky-subtracted images
for each filter to estimate the sky background after removal of the faintest
sources. The stacked images represent the weighted median of the individual
images combined with the Lanczos3 interpolation kernel. The reference
coordinate system is the FK5 or J2000 system with the distorted tangential 
projection type
\item Sky subtraction is made in two steps and created from a sample of
30 images before subtracting the sky from the detrended images
\item In some cases sky masking has to be implemented in addition to
the normal procedure if it turned out to be impossible to estimate the
sky level for some pixels
\item Assessment of the quality of sky-subtracted images and catalogues.
After visual inspections, a few images ($<$ 20) had to be rejected due to
luminosity gradients, cosmetics defects or large scale patterns
\end{enumerate}

The final step consists of generating sky--subtracted images to compute
the astrometric and relative photometric calibration with field to
field rescaling. Individual stacks were created for the 18 images
in methane OFF and ON as well as two mosaics containing the nine pointings
of each filter. The catalogues, generated with SExtractor \citep{bertin96},
contain the pixel and world coordinates, aperture fluxes and magnitudes with 
their associated errors as well as quality flags. The final merged catalogue 
with the methane ON and OFF data contains 166,265 sources, of these 131,937 
have quality flags equal to zero indicating that they are robust detections.
There are 92,918 sources with photometric
error bars less than 0.3 mag in the methane colour. The completeness limit 
of the methane survey is 20.3 and 20.2 mag in the CH$_{\rm 4\,off}$ and 
CH$_{\rm 4\,on}$, respectively. These limits are deduced from the points
where the histograms deviate from a powerlaw fit to the counts.

\subsection{Deep UKIRT WFCAM $ZYJ$ survey}
\label{deepUsco:phot_obs_WFCAM}
\subsubsection{WFCAM Observations}

The UKIRT wide--field camera (WFCAM) is equipped with four 2048$\times$2048 
Rockwell detectors spaced by 94\% and sensitive to near-infrared (1.0--2.5 
microns) wavelengths \citep{casali07}. The pixel size is 0.4 arcsec, giving 
a coverage of 0.19 square degree in a single exposure (called pawprint). 
Four pawprints are required to fill in the gaps between the detectors and 
cover a filled square of 0.75 square degree (or tile). The camera is 
equipped with several filters, including the $Z$ and $Y$ filters specific 
to WFCAM as well as the standard Mauna Kea Observatory \citep{tokunaga02} 
$JHK$ filters \citep{hewett06}.

\begin{figure}
  \centering
  \includegraphics[width=\linewidth, angle=0]{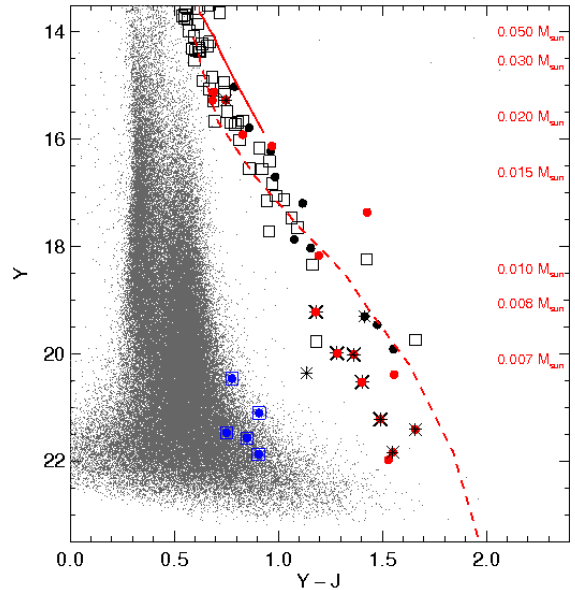}
  \caption{($Y-J$,$Y$) colour-magnitude diagram for $\sim$1.7 square degree
in USco. The small dots are all sources detected along the line of sight of 
the association. Open squares are spectroscopically confirmed members of the 
USco association from the UKIDSS GCS \citep{lodieu07a,lodieu08a}. Black 
filled circles are photometric candidates identified in the deep $YJ$ survey
alone, the red circles being new ones. Asterisks are later classified as
photometric and/or spectroscopic non members using additional datasets
presented in this paper. Blue circles with blue open squares
are the five methane candidates undetected in the optical and with $Y,J$ 
photometry. Overplotted are the NextGen \citep[solid line;][]{baraffe98} and 
DUSTY \citep[dashed line;][]{chabrier00c} 5 Myr isochrones shifted at a 
distance of 145 pc.
}
  \label{fig_deepUsco:cmd_yjy}
\end{figure}

We have conducted a deep $Y,J$ imaging survey with WFCAM of the central
region in the USco previously surveyed by the UKIRT Infrared Deep Sky
Survey (UKIDSS) Galactic Clusters Survey (GCS) \citep{lodieu07a}. The 
observations took place in May and June 2008 
(Table~\ref{tab_deepUsco:log_WFCAM}).
Weather conditions were generally good with clear skies and seeing better
than 1.0 arcsec as measured on the images. We have obtained eight pawprints 
in the $Y$ and $J$ filter, covering a total of $\sim$1.7 square degrees. An 
additional $Z$--band image was obtained for four pawprints because of the 
high airmass of USco from Mauna Kea and our seeing requirements which
hampered achieving regularly seeing better than 1.2 arcsec in this filter.
The central coordinates of each pawprint are provided in 
Table~\ref{tab_deepUsco:log_WFCAM}. The on--source individual exposure times
were set to 22.25, 20, and 10 seconds in the $Z$, $Y$, and $J$ filters,
respectively, repeated twice in $Z,Y$ and three times in $J$.
We used a 2$\times$2 microstepping pattern and a 5 point dither 
arrangement with offsets of 3.2 arcsec to subtract the sky properly.
We reached depths of 5$\sigma$ at 22.0--22.3 mag and 21.5--21.7 mag 
in $Y$ and $J$, respectively. The resulting ($Y-J$,$Y$) colour--magnitude 
diagram is displayed in Figure~\ref{fig_deepUsco:cmd_yjy}. We observe a
gap between potential members and field stars as previously reported in our 
studies of USco \citep{lodieu06,lodieu07a} and the Pleiades
\citep{lodieu07c}. However, as we discuss in the next sections, most of
the new photometric candidates identified in this deep survey which
follow the sequence of USco spectroscopic members turn out to be non-members.

%
%
%
\begin{table}
 \centering
 \caption[]{Log of the UKIRT WFCAM photometric observations.}
 \begin{tabular}{@{\hspace{0mm}}c c c c c c@{\hspace{0mm}}}
 \hline
 \hline
Tile &   R.A.   &     dec     &  Repeat         &  Date     & Seeing \\
 \hline
\#1  & 16:06:00 & $-$23:09:00 & 4$\times$$J$ & 03May2008 &  1.0$''$ \\
\#1  &          &             & 1$\times$$J$ & 04May2008 &  $\sim$0.9$''$ \\
\#1  &          &             & 4$\times$$Y$ & 04May2008 &  $\sim$0.9$''$ \\
\#1  &          &             & 1$\times$$Z$ & 03Jun2008 &  $\sim$1.0$''$ \\
\hline
\#2  & 16:13:12 & $-$22:15:00 & 5$\times$$J$ & 05May2008 &  0.5--1.0$''$ \\
\#2  &          &             & 2$\times$$Y$ & 30May2008 &  $\sim$1.0$''$ \\
\#2  &          &             & 2$\times$$Y$ & 31May2008 &  $\sim$0.7$''$ \\
  \hline
 \label{tab_deepUsco:log_WFCAM}
 \end{tabular}
\end{table}
%

%
%
%
\subsubsection{Data reduction}

All data taken with WFCAM were reduced automatically by the Cambridge 
Astronomy Survey Unit (CASU)\footnote{http://casu.ast.cam.ac.uk}
even if there are not part of the UKIDSS project \citep{lawrence07}.
The description of the automatic pipeline data reduction is explained
in \citet{irwin04} with additional updates detailed on the CASU 
webpage\footnote{http://casu.ast.cam.ac.uk/surveys-projects/wfcam/technical}.
The processing involve several technical steps before generating the final
photometric catalogues, including flat fielding,
sky subtraction, correction for field distortion, removal of low level 
pseudo-periodic ripples in dark images (or decurtaining), and removal of 
persistence and crosstalk artifacts due to the presence and reminiscence
of bright stars on infrared detectors.
The catalogues can then be retrieved from the WFCAM Science Archive
\citep[WSA;][]{hambly08} to extract the astrometry and photometry.

As a consequence of multiple repeats of each pawprint for each filter, 
an additional step was necessary to combine all images of each pawprint 
in each filter.
At WFAU (Wide Field Astronomy Unit) pawprints from each minimum scheduled 
block (MSB) were first subjected to quality control using an automated set 
of procedures, and those that were deemed acceptable were used in all 
further curation and released and the others were deprecated 
\citep[see][for details]{collins09}. The quality controlled pawprints were 
then grouped together, using the metadata in the WSA tables. The grouping 
occurred in two stages: first by position to find all unique pointings 
(where pointings were within 4 arcmin of each other) and then grouped by 
pointing, filter and level of microstepping to determine deep products. 
If a particular pointing and filter combination contains more than one 
level of microstepping, then the level of microstepping that corresponds 
to data with the greatest total exposure time is selected on a 
product-by-product basis. The stacking code used can only deal with one 
level of microstepping, so this is a necessity. 

Deep pawprint stacks and confidence images were then created by looping 
through the list of stacks and selecting pawprint science frames (the stacks 
created from processing each MSB) that match the pointing, filter and 
microstep criteria. The CASU code ``fitsio\_stack'' was used for the stacking, 
and then a catalogue was created for each stack using the codes 
``cir\_imcore'', ``cir\_classify'' and ``cir\_catcoord'', all produced at 
CASU (see Irwin et al.\ in prep). 
The new deep stacks and catalogue data were then ingested into the WSA\@. 

The merged passband source table was produced from the deepest data in 
each pointing/filter, i.e.~from the deep stack catalogues. \citet{hambly08} 
provides details on the production and contents of the Source table. 
Multi--epoch tables were produced as described in \citet{cross09b}.
The final combined catalogues could then be retrieved directly
with a Structured Query Language (SQL) query sent through the WSA webpage.

\subsection{INT WFC optical $z$-band imaging}
\label{deepUsco:phot_obs_INT}
\subsubsection{INT Observations}

The Issac Newton Telescope (INT) is a 2.54-m optical telescope currently 
equipped with two instruments, one of which being a wide--field camera (WFC)
located at its prime focus. The WFC consists of four thinned EEV 
2048$\times$x4096 Charge-Coupled Devices (CCDs), offering a field-of-view of 
approximately 34 arcmin and a pixel size of 0.33 arcsec. The inter-spacing
between the CCDs is small, resulting in a loss of coverage of the order of
1 arcmin.

We carried out a $z$-band imaging survey of the full coverage of the
GCS Science Verification phase presented in \citet{lodieu07a} over several
nights on 19--25 April 2006\@. In this paper, we mainly focus on the
area overlapping with the methane and deep $YJ$ survey to complement the 
near-infrared photometry and add constraint on the membership of potential
cluster member candidates. The nights of 19 and 25 April offered excellent 
weather with good seeing and stable conditions. The weather conditions on 
20 April were also good but the seeing was variable. The night of 21 April 
was completely lost due to high humidity. The nights on 22 and 23 April 
were windy and the night on 24 April had humidity of 80\%.

\subsubsection{Data reduction}

The WFC data were reduced using the Cambridge Astronomical Survey Unit CCD 
reduction toolkit \citep{irwin01} to follow standard steps, namely, 
subtraction of the bias, non-linearity corrections, flat-fielding and 
astrometric calibration. However, in addition, each frame was corrected for 
fringing using a fringe map constructed from a clipped stack of the images 
from all our pointings. Subsequently, photometry was performed on the 
reduced images using a circular window with a diameter of 1.5$\times$ the 
full width half maximum of the mean point spread function. Finally, we created 
catalogues of all the sources detected in the frames and morphologically 
classified these as either stellar or non-stellar. We haven't calibrated 
photometrically the INT images with a specific standard star but instead 
used the WFCAM $Z$-band from the Science Verification phase to estimate 
the depth of the INT observations. We estimate the 3$\sigma$ depths to 
$Z$=20.7--21.1 mag on the 19, 23--25 April 2008 while images taken on 20 and 
22 April are shallower with depths of 20.0$\pm$0.2 and 19.5$\pm$0.3, 
respectively. Thus, the INT images are usually deeper than the UKIDSS GCS
survey, except for the observations on 20 and 22 April 2008\@. Note that 
these depths are given in the WFCAM $Z$ filter and not the INT $z$ filter.

\subsection{Spectroscopic observations}
\label{deepUsco:phot_obs_spectro}
\subsubsection{X-shooter observations}
\label{deepUsco:XSHOOTERspec_obs}

We carried out optical to near-infrared spectroscopy with the UV- to 
$K$-band intermediate resolution (R$\sim$3500), high efficiency spectrograph 
X-shooter \citep{dOdorico06} mounted on the Cassegrain focus of the Very 
Large Telescope (VLT) Unit 2\@. Observations were carried out over several 
nights during period 85 (April to September 2010) in service mode by the 
staff at Paranal, Chile. The log of the observations is detailed in
Table~\ref{tab_deepUSco:log_XSHOOTER}. Weather conditions met
our initial requirements: no constraints on the moon, seeing better
than 1~arcsec (satisfied in most cases), and thin cirrus acceptable.

X~Shooter is a multi wavelength cross--dispersed echelle spectrograph made 
of three independent arms covering simultaneously the ultraviolet (UVB; 
0.3--0.56 microns), visible (VIS; 0.56--1.02 microns), and near--infrared 
(NIR; 1.02--2.48 microns) wavelength ranges thanks to the presence of two 
dichroics splitting the light. The spectrograph is equipped with three
detectors: a 4096$\times$2048 E2V CCD44-82, a 4096$\times$2048 MIT/LL 
CCID\,20, and a 2096$\times$2096 Hawaii 2RG for the UVB, VIS, and NIR arms, 
respectively. We used the 1.5 arcsec slit (1.6 arcsec for the UVB arm) 
to achieve a nominal resolution of 3300 (9.9 pixel per 
full--width--half--maximum) in the UVB and NIR arms and 5400 (6.6 pixel per 
full--width--half--maximum) in the VIS arm.

All observations were carried out at parallactic angles. The X--Shooter 
acquisition is made through an optical CCD, implying that our targets
are not visible on the images obtained by the acquisition camera screen. 
Therefore, we used nearby reference 
stars to put our targets on the slit. Unfortunately, due to an error in
the offset computation, one candidate, USco J160812.99$-$230431.43, 
observed in April 2010 was out of the slit and we do not have spectra 
for it.
The exposure times were scaled according to the brightness of the target
(see Table~\ref{tab_deepUSco:log_XSHOOTER}) to achieve a minimum 
signal--to--noise of 20 after degrading the resolution by a factor of 5\@.
Calibrations were taken according to the instrument calibration plan,
including bias, dark, flat field and arc frames.
The spectra of four confirmed photometric candidates, normalised at 1.265 
microns, are displayed in Figure~\ref{fig_deepUsco:final_NIR_spectra}.

%
%
%
\begin{table*}
 \centering
 \caption[]{Log of the X-Shooter spectroscopic observations.
            Note that USco J160835.54$-$225311.4 was observed
            three times. Note that USco J160812.99$-$230431.43,
            not listed in this table, was out of the slit due 
            to an error in the offset computation.}
 \begin{tabular}{@{\hspace{0mm}}c c c c c c c c c@{\hspace{0mm}}}
 \hline
 \hline
ID & R.A.        &     dec      &   $J$  &  Date & ExpT & Seeing & Airmass & Reference star \\
 \hline
 4 & 16:07:55.42 & $-$22:33:52.52 & 18.039 &  24Apr2010 & 20min & 0.60--0.69 & 1.010 & 16:07:56.07 $-$22:33:24.6 ($I$=10.60 mag) \\ 
 7 & 16:08:42.38 & $-$22:23:25.86 & 18.649 &  29Jul2010 & 45min & 0.64--0.67 & 1.087 & 16:08:43.92 $-$22:23:29.4 ($I$=14.47 mag) \\ 
 3 & 16:07:16.98 & $-$22:42:14.06 & 19.730 &  11Jun2010 & 60min & 0.66-0-.68 & 1.212 & 16:07:14.41 $-$22:41:59.8 ($I$=15.45 mag) \\ 
 8 & 16:09:00.40 & $-$23:11:50.95 & 19.751 &  26Jul2010 & 60min & 1.14--1.26 & 1.122 & 16:09:02.01 $-$23:12:04.11($I$=14.06 mag) \\ 
 6 & 16:08:35.54 & $-$22:53:11.37 & 20.284 &  19Aug2010 & 40min & 1.25--1.30 & 1.038 & 16:08:38.72 $-$22:53:04.0 ($I$=15.09 mag) \\ 
 6 & 16:08:35.54 & $-$22:53:11.37 & 20.284 &  19Aug2010 & 40min & 1.25--1.30 & 1.111 & 16:08:38.72 $-$22:53:04.0 ($I$=15.09 mag) \\ 
 6 & 16:08:35.54 & $-$22:53:11.37 & 20.284 &  20Aug2010 & 40min & 0.85--0.96 & 1.056 & 16:08:38.72 $-$22:53:04.0 ($I$=15.09 mag) \\ 
  \hline
 \label{tab_deepUSco:log_XSHOOTER}
 \end{tabular}
\end{table*}
\subsubsection{Data reduction}
\label{deepUsco:XSHOOTERspec_DR}

We reduced the X--Shooter spectra on each arm independently using version
1.3 of the pipeline and following the instructions described in the manual 
\citep[(Issue 4; 30 november 2010)][]{goldoni06,modigliani10}.
None of the five targets have flux in the ultraviolet arm (UVB) so we do 
not discuss these data further. 

We followed the steps enumerated in the manual using the physical 
model option to construct a final 2D spectrum
for each target and its respective standard star both in the visible (VIS)
and infrared (NIR) arm. First, we created a bad pixel map for the NIR
detector using a set of 40 linearity frames aimed at identifying non-linear
pixels. This step is not required for the VIS arm. Next, we 
created a master bias and master dark for the VIS and NIR arm, respectively.
Afterwards, we determined a first guess order and then we refined
the line tables by illuminating the X--Shooter pinhole with a continuum
lamp. Later, we created a master flat and an order table tracing
the flat edges before establishing the two--dimensional map of the
instrument. Subsequently, we determined the efficiency of the whole 
system made of the telescope, instrument, and detector. Finally, we
generated a 2D spectrum in stare mode for the target and its 
associated standard, both in the VIS and NIR arms. 

The next steps were carried out under the IRAF environment\footnote{IRAF is
distributed by National Optical Astronomy Observatory, which is operated by 
the Association of Universities for Research in Astronomy, Inc., under 
contract with the National Science Foundation.} although an optimal 
extraction option is available in the pipeline's release. First, we
extracted a 1D spectrum from the 2D images of the VIS and NIR arm for the 
target and the star star using the IRAF routine {\tt{APSUM}}. Only few 
spectrophotometric standard stars have been been fully calibrated from the
UV to the near-infrared spectral range at the
time of writing. For this reason, all of our targets are associated with
the GD\,153 white dwarf standard, except candidate number 6 which was
observed just before another white dwarf standard, Feige 110\@. We
downloaded the calibrated spectra of GD\,153 and Feige 110 originating
from observations conducted with the Hubble Space Telescope and covering 
the UVB, VIS, and NIR arms\footnote{Ascii files containing fluxes from the 
ultraviolet to the near-infrared can be downloaded from
http://www.stsci.edu/hst/observatory/cdbs/calspec.html}.
Afterwards, we divided the 1D spectrum of the standard star from the 
1D spectrum of the science target and multiplied by the Hubble spectrum of 
GD\,153 and Feige 110 for candidates 1 to 5 and candidate, respectively.

The X--Shooter spectra are contaminated by the presence of sky lines.
We attribute this to the current version (Issue 1.3) of the pipeline which
is not yet optimised to minimise the residuals of the numerous sky lines
present over the full wavelength range covered by X--Shooter, particularly
in the stare mode reduction. The near--infrared (1.0--2.25 microns) 
spectra of four photometric candidates resembling L--types are displayed in 
Figure~\ref{fig_deepUsco:final_NIR_spectra}.
The optical spectra of USco J160842.38$-$222325.86 and
USco J160716.98$-$2242.1406 are shown in 
Figure~\ref{fig_deepUsco:final_VIS_spectra}; the other two objects
exhibiting no flux blueward of 1$\mu$m.
The spectrum of USco J160900.40$-$231150.95 classifies it as a quasar
at a redshift of 0.8789 and is discussed in more detail in 
Section~\ref{deepUsco:XSHOOTERspec_QSO}.

%
%
\section{Search for cluster T--type candidates}
\label{deepUsco:select_Tdwarfs}

A search for T--type candidate members of the USco Association is likely easier 
than for the Trapezium Cluster and the Pleiades due to a combination of it's age 
and distance. Indeed, the apparent magnitude of a 3 M$_{\rm Jup}$ T--type 
(effective temperature below $\sim$1100\,K) at 5 Myr and 145 pc is predicted 
to be $\sim$0.5 and $\sim$4.5 mag brighter in $J$ than in the Trapezium Cluster 
and the Pleiades, respectively, based on the COND models for dust--free brown 
dwarfs \citep{baraffe02}. Furthermore, USco exhibit a significant proper 
motion \citep[$-$11,$-$25 mas/yr][]{deZeeuw99} so that the membership of any 
T--type candidate could be confirmed (or otherwise) with 
a 3 to 5 year baseline. Finally, star formation has ended in USco 
\citep{walter94} and the region we have focused our deep survey on has low 
foreground extinction \citep[Av $\leq$ 2 mag;][]{preibisch98}.

In this section, we describe our photometric selection of cluster T--types 
starting from the methane imaging alone. Then, we look at their optical 
photometry using the INT data. Afterwards, we investigate their $Y,J$ 
photometry to check if these are consistent with T--type members of USco.

The absolute $J$ magnitudes of old ($>$1 Gyr) field T--type brown dwarfs are 
fainter than M$_{J}$ $\sim$ 13.5--14.0 mag 
\citep{vrba04,liu07,leggett10a,marocco10}, 
corresponding to an apparent magnitude of $J$ $\sim$ 19.3--19.8 mag at the 
distance of USco. The $H$-band absolute magnitude is roughly 0.5 mag 
brighter. Therefore, we proceed as follows to select potential ``methane'' 
candidates. We 
considered only sources with error bars on their methane colour less than 
0.3 mag. First, we computed the median values of the methane colour,
per 0.5 mag interval of the methane OFF band from 17.5 to 21 mag 
(Table~\ref{tab_deepUSco:CH4_stats}). We set the brightness limit to 17.5 mag 
to account for the fact that substellar objects are brighter when younger, 
and the blue $J-H$ and $H-CH_{\rm 4off}$ trends exhibited towards lower 
temperatures -- for example, there is 
a difference of about one magnitude for a 5 Myr--old 1000\,K brown dwarf 
compared to a 1 Gyr T--type brown dwarf at the same effective temperature 
\citep{burrows97,chabrier97}. Then, we calculated the median absolute 
deviation (MAD) i.e.~the median of the set of absolute values of 
deviation from the central value computed as above. 
Finally, we selected as potential candidates all those objects lying
3$\sigma$ to the blue of the median methane value in each bin where $\sigma$ 
was robustly computed as 1.48$\times$MAD. The median and $\sigma$
values are reported in Table~\ref{tab_deepUSco:CH4_stats} along with the
number of candidates.

%
%
%
\begin{table}
 \centering
 \caption[]{Statistical selection of potential candidates in
            USco from the methane survey. `Median' is the
            median value of the methane colour in each 
            magnitude interval; STD is the standard deviation,
            `Nb' is the number of objects used to compute the
            median and standard deviation, and the last column
            represents the number of candidates
            satisfying the 3$\sigma$ methane selection.}
 \begin{tabular}{@{\hspace{0mm}}c c c c c c@{\hspace{0mm}}}
 \hline
 \hline
$J$ range & Median &  $\sigma$  &   Nb  &  Candidates \cr
 \hline
17.5--18  & -0.060 & 0.058 &  3478 &   863  \cr
18--18.5  & -0.079 & 0.080 &  4497 &  1088  \cr
18.5--19  & -0.109 & 0.114 &  6027 &  1398  \cr
19--19.5  & -0.148 & 0.155 &  8226 &  1846  \cr
19.5--20  & -0.209 & 0.217 &  9943 &  1934  \cr
20--20.5  & -0.288 & 0.290 & 10452 &   870  \cr
20.5--21  &  0.338 & 0.339 &  7121 &    19  \cr
\hline
17.5--21  &   ---  &  ---  & 49744 &  8018  \cr
  \hline
 \label{tab_deepUSco:CH4_stats}
 \end{tabular}
\end{table}
To eliminate contaminants, we cross correlated this large list of candidates 
with the INT optical photometric catalogue using a matching radius of 
2 arcsec. We found 5410 sources with optical photometry out of the 8018 
methane candidates, implying that 2608 have no detected optical counterpart. 
About 65\% of those candidates without optical counterpart down to the
depth of the INT survey are covered by the deep $YJ$ data, leaving about 
1750 sources to investigate further. Among these $\sim$1750 sources, 806 have 
a counterpart within 2 arcsec in the deep $YJ$ survey.

We investigated the near--infrared photometry of these 806 candidates
to reject contaminants. We computed the $J-CH_{\rm 4off}$ colours
for each candidate and considered only sources bluer than 0.5 mag in
$J-CH_{\rm 4off}$ \citep{tinney05,goldman10,penya11a}. 
Applying this colour cut leaves one object with a negative
$J-CH_{\rm 4off}$ colour, one object with $J-CH_{\rm 4off}$ $\leq$ 0.2 
mag, 4 candidates with $J-CH_{\rm 4off}$ = 0.25-0.30 mag, 
12 candidates with $J-CH_{\rm 4off}$ = 0.3-0.4 mag,
and 26 candidates with $J-CH_{\rm 4off}$ = 0.4-0.5 mag.
However, we discarded the candidate with the bluest $Y-J$ colour because it
turned out to be a mismatch. Additionally, we rejected 13 of these sources
as they were detected in the WFCAM $Z$ imaging and a further 29 with $Y-J$
colours that are too blue ($Y-J$ $\leq$ 0.7 mag; including 4 already 
detected on the WFCAM $Z$ images) to be consistent with 
T-type members. Thus there are 5 candidates which remain of interest to us.
We checked those five sources in the WFCAM $YJ$ (no $Z$-band images
available) and the WIRCam methane images to detect any false positive but 
did not find any. The bluest object in $J-CH_{\rm 4off}$ and thus the most 
interesting candidate, has a methane colour of $-$0.87 mag and 
$Y-J$ = 0.85$\pm$0.10\@. The coordinates and photometry of these candidates 
is provided in Table~\ref{tab_deepUSco:CH4_candidates}.

We looked at the images of the recent data release of the Wide-Field
Infrared Survey Explorer \citep[WISE;][]{wright10} to check whether these
T--type candidates are detected at 3.4 and 4.6 microns. None of the five
candidates are detected. However, non--detections do not constitute a strong 
constraint on their spectral type since T--types exhibit $J-3.4$ and $J-4.6$
colours in the 2--4.5 and 0--3 mag range 
\citep{patten06,leggett10a,mainzer11} 
which does not guarantee a mid-infrared detection at to the WISE 
sensitivities (16.5 and 15.5 mag at 3.4 and 4.6 microns, respectively).

%
%
%
\begin{figure}
  \centering
  \includegraphics[width=\linewidth, angle=0]{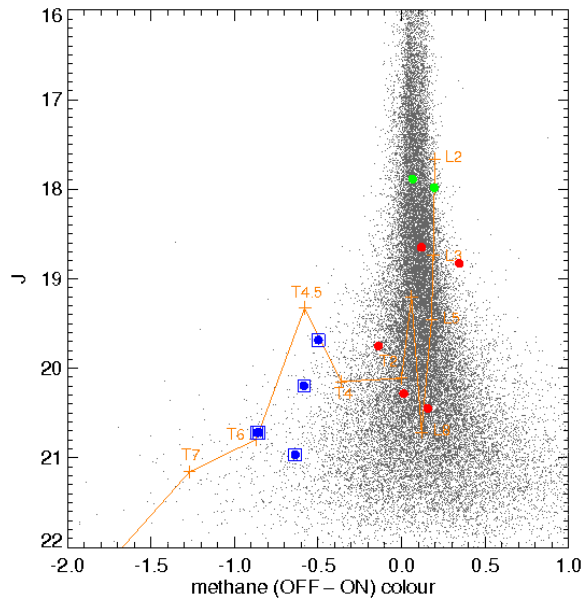}
  \caption{(CH$_{\rm 4off}$ $-$ CH$_{\rm 4on}$,$J$) colour-magnitude diagram 
for $\sim$0.95 square degree in USco. The small grey dots are all sources
detected in the deep $YJ$ survey with methane photometry. Green symbols are
USco photometric candidates with a near-infrared spectrum: the bluer object
is a field L2 dwarf whereas the redder source is L1 dwarf member of USco.
Red dots are new candidates selected from the deep $YJ$ survey for which
methane photometry is available. Overplotted are synthetic colours for
spectral types L2, L3, L5, L8, T0, T2, T4, T4.5, T6, T7, and T8
\citep[orange lines with plus signs to mark the spectral types;][]{tinney05}.
Blue circles surrounded by open squares are the five methane candidates 
undetected on the INT images and with $Y,J$ counterparts.
}
  \label{fig_deepUsco:cmd_CH4J}
\end{figure}
%

%
%
%
%
\begin{table}
 \centering
 \caption[]{Photometry for the methane T--type candidates without optical
            counterparts and detected in the deep $YJ$ survey. The last
            column gives a priority (P) of interest, 1 being the highest.}
 \begin{tabular}{@{\hspace{0mm}}c @{\hspace{2mm}}c c @{\hspace{2mm}}c c c @{\hspace{2mm}}c@{\hspace{0mm}}}
 \hline
 \hline
R.A.        &     dec      &  $Y_{\rm deep}$  &  $J_{\rm deep}$   & CH$_{\rm 4off}$ & CH$_{\rm 4on}$ & P \cr
 \hline
16:08:35.98 & $-$22:29:11.1 & 21.869 & 20.964 & 20.471 & 21.108 &  3  \cr
16:08:45.73 & $-$22:29:53.5 & 21.469 & 20.718 & 20.290 & 21.145 &  3  \cr
16:08:47.80 & $-$22:29:04.5 & 21.568 & 20.718 & 20.585 & 21.450 &  1  \cr
16:09:55.91 & $-$22:33:45.7 & 21.103 & 20.194 & 19.811 & 20.395 &  2  \cr
16:10:04.76 & $-$22:32:30.6 & 20.462 & 19.685 & 19.257 & 19.754 &  3  \cr
  \hline
 \label{tab_deepUSco:CH4_candidates}
 \end{tabular}
\end{table}
%

%
%
\section{Bridging the gap between M and T--types}
\label{deepUsco:select_Ldwarfs}

In this section we describe the photometric selection employed to identify
further USco member candidates in the area common to the deep $YJ$ survey and
methane imaging. Additional membership constraints are derived from the
infrared $Z$ photometry as well as crossmatching with the UKIDSS 
GCS \citep{lodieu07a} and its associated spectroscopic follow--up
\citep{lodieu08a,lodieu11a}.

We found that the USco sequence is well separated from field stars in 
several colour--magnitude diagrams (including the ($Y-J$,$Y$) diagram) 
from the central region surveyed during the Science Verification (SV) phase
of the UKIDSS GCS \citep{lodieu07a}. The cluster 
sequence extends down to 10 M$_{\rm Jup}$ and possibly below depending on 
the validity of the mass estimates provided by current theoretical models
\citep{baraffe02}. We note that we clearly detect the M7/M8 gap at 
$Z$ $\sim$ 15.5 mag and $J$ $\sim$ 14 mag proposed to by \citet{dobbie02b}. 

We started off our photometric selection using the ($Y-J$,$Y$) 
colour--magnitude diagram (Figure~\ref{fig_deepUsco:cmd_yjy}). We used 
the sample of spectroscopic members extracted from the UKIDSS GCS 
\citep{lodieu08a} to define our selection criteria. The sequence appearing 
in the deep ($Y-J$,$Y$) colour--magnitude diagram matches well the sequence
previously seen in the GCS SV data \citep{lodieu07a} after applying the
offset of $+$0.08 mag to the SV data (see ``Known Issues'' for UKIDSS Data
Release 3 in the WSA). We queried the WSA to extract coordinates and
photometry of all point sources with good quality fainter than $J$ = 10.5 mag 
(saturation level of the deep survey). This catalogue contains 172,209 
objects. However, we focused our search only on sources fainter than
$Y$ = 15 mag because we aim to uncover USco members with
masses below $\sim$30 M$_{\rm Jup}$ and, more specifically, 5 Myr--old
L-- and T--types. Similarly, we considered only sources brighter
than the 5$\sigma$ detection limit in $Y$ i.e.~$Y$ $\leq$ 22.2 mag. As a 
consequence, we picked out all objects to the right of a line extending 
from ($Y-J$,$Y$) = (0.65,15.0) to (1.50,22.0) in the ($Y-J$,$Y$) diagram.
This selection returned 30 sources detected in $Y$ and $J$, including
three with $Z$-band photometry. Among these 30 sources, four have been 
removed from the final candidate list because they were either residual crosstalk 
artifacts (three objects) or at the edge of the detector (one object). Thus,
we are left with 26 reliable candidates which warrant further investigation.

We crossmatched these 26 candidates against the list of members
identified in the GCS SV, resulting in ten common sources
(Table~\ref{tab_deepUsco:tab_candYJ_SV2}). We have recovered all candidates 
from our GCS SV work in our selection. Eight of them are known spectroscopic 
members with spectral types ranging from M8 to L1 \citep{lodieu08a}, one is 
classified as a spectroscopic non member 
\citep[USco J160956.34$-$222245.5; dL2;][]{lodieu08a}, and the remaining 
one has an optical spectral type (USco J161154.39$-$223649.3; M6.25) from
our multi-fibre spectroscopic follow-up \citep{lodieu11a}.
We note that the new $Y-J$ colour (1.55$\pm$0.06 mag) of 
USco J160843.43$-$224516.1 differs from the old measurement from the GCS 
SV ($Y-J$ = 1.18$\pm$0.20), putting this object back on the cluster 
sequence in the ($Y-J$,$Y$) diagram. We conclude that this spectroscopic 
member may be variable and/or may possess circumstellar material.

%
%
%
%
\begin{table}
 \centering
 \caption[]{Coordinates (in J2000), near-infrared $YJ$ magnitudes and
            their associated photometric error bars for the 10 candidates
            extracted from our deep $YJ$ survey and already identified
            in the GCS Science Verification survey \citep{lodieu07a}.
            The last column provides the spectral types derived from
            spectroscopy \citep{lodieu08a,lodieu11a}.
            }
 \begin{tabular}{@{\hspace{0mm}}c @{\hspace{2mm}}c @{\hspace{2mm}}c @{\hspace{2mm}}c c@{\hspace{0mm}}}
 \hline
 \hline
R.A.        &     dec       &  $Y \pm errY$    &  $J \pm errJ$    & SpT \cr
 \hline
16:06:48.18 & $-$22:30:40.2 & 15.787$\pm$0.003 & 14.928$\pm$0.002 & M8/M8.5 \cr
16:07:27.83 & $-$22:39:04.0 & 18.030$\pm$0.008 & 16.873$\pm$0.006 & L1 \cr
16:07:37.98 & $-$22:42:47.0 & 17.865$\pm$0.008 & 16.788$\pm$0.006 & L0 \cr
16:08:18.43 & $-$22:32:25.1 & 17.194$\pm$0.005 & 16.077$\pm$0.004 & L0 \cr
16:08:43.43 & $-$22:45:16.1 & 19.912$\pm$0.030 & 18.358$\pm$0.015 & L1 \cr  
16:08:47.45 & $-$22:35:47.9 & 16.701$\pm$0.004 & 15.716$\pm$0.004 & M9 \cr
16:09:18.68 & $-$22:29:23.8 & 19.457$\pm$0.020 & 17.981$\pm$0.012 & L1 \cr
16:09:56.34 & $-$22:22:45.6 & 19.302$\pm$0.019 & 17.887$\pm$0.011 & dL2 (NM) \cr
16:10:47.14 & $-$22:39:49.5 & 16.218$\pm$0.003 & 15.254$\pm$0.003 & M9/M8.5 \cr
16:11:54.39 & $-$22:36:49.3 & 15.015$\pm$0.002 & 14.225$\pm$0.002 & M6.25  \cr
  \hline
 \label{tab_deepUsco:tab_candYJ_SV2}
 \end{tabular}
\end{table}
%

%
%
%
\begin{figure*}
  \centering
  \includegraphics[width=0.47\linewidth, angle=0]{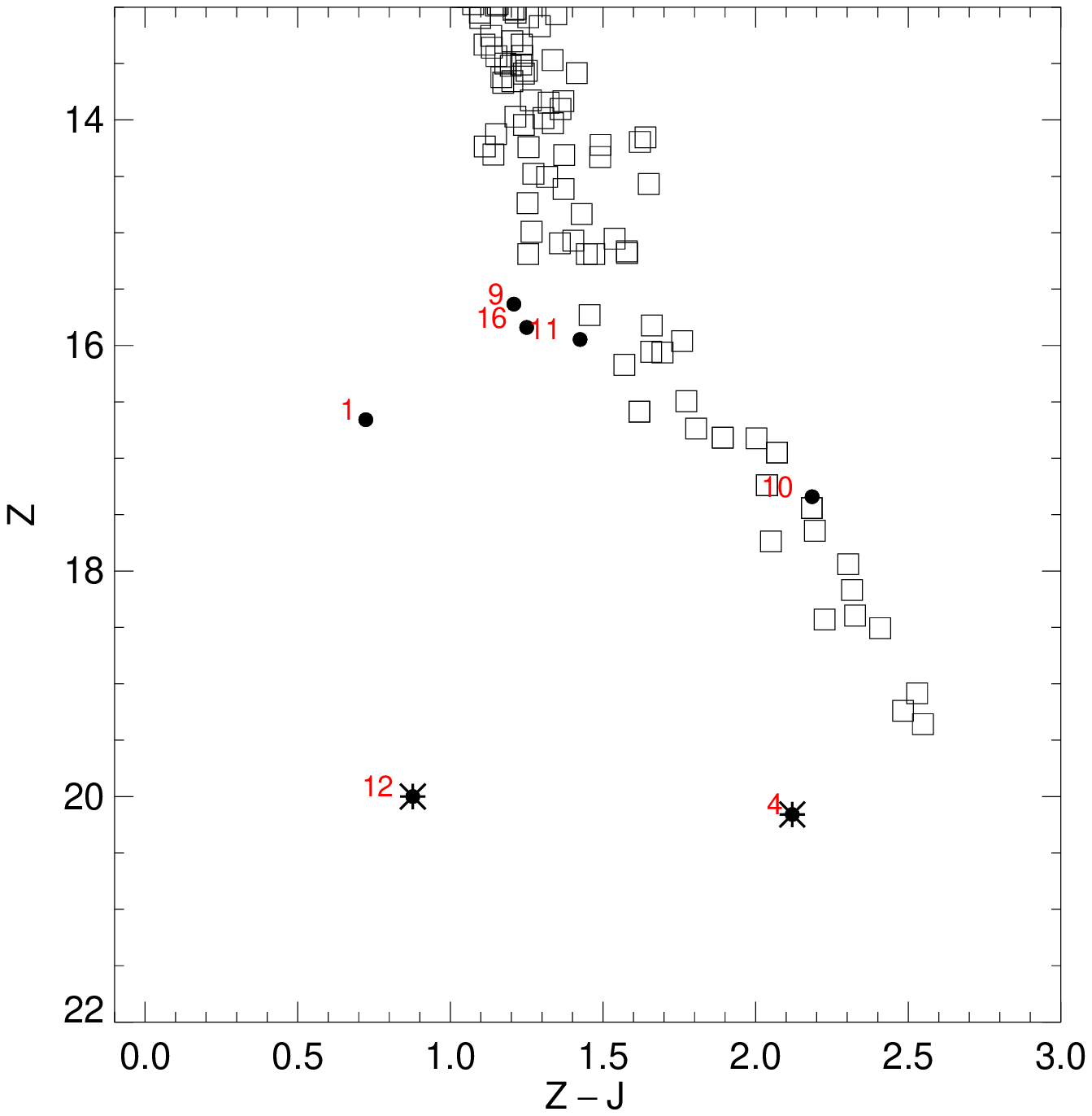}
  \includegraphics[width=0.47\linewidth, angle=0]{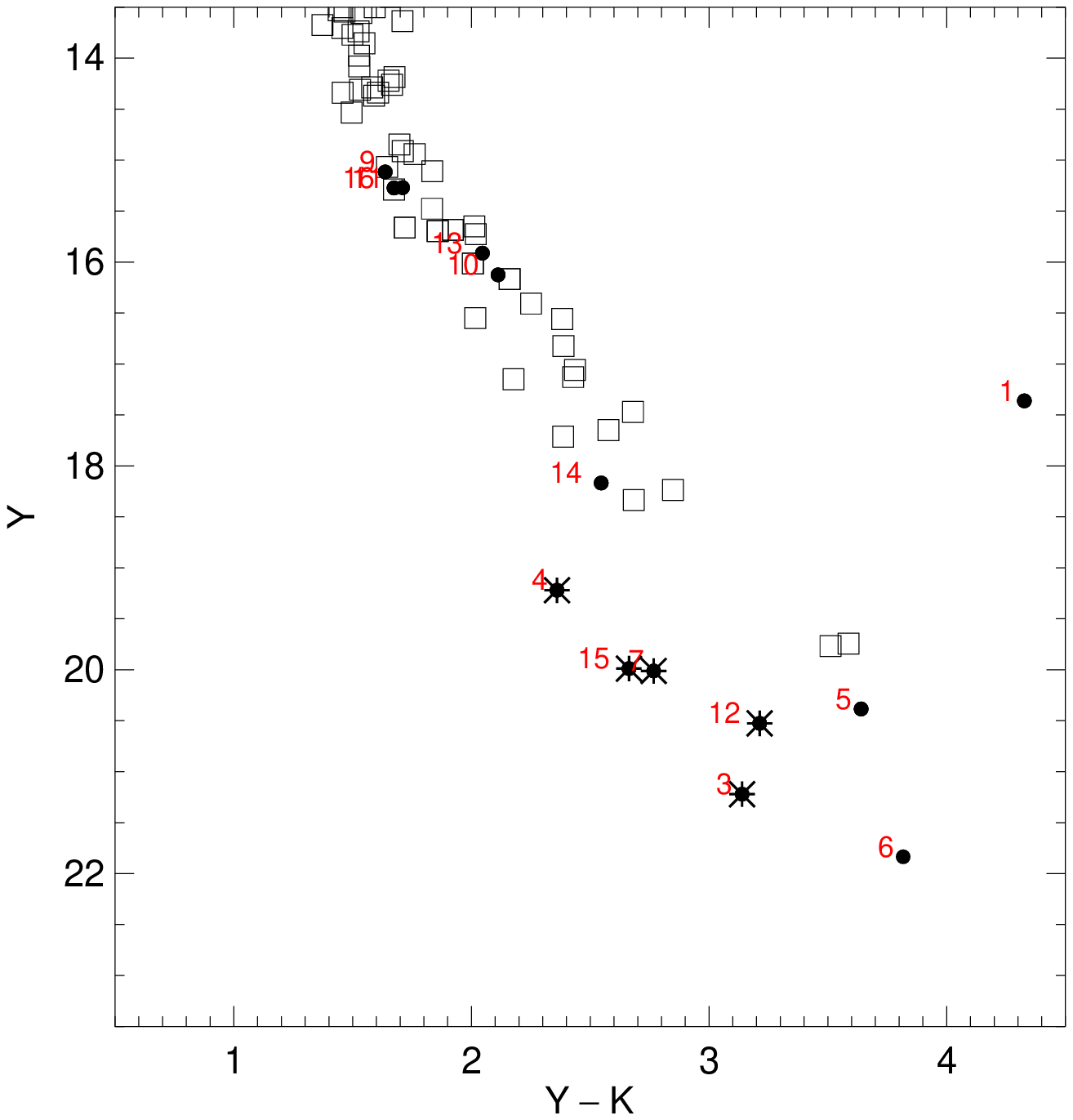}
  \includegraphics[width=0.47\linewidth, angle=0]{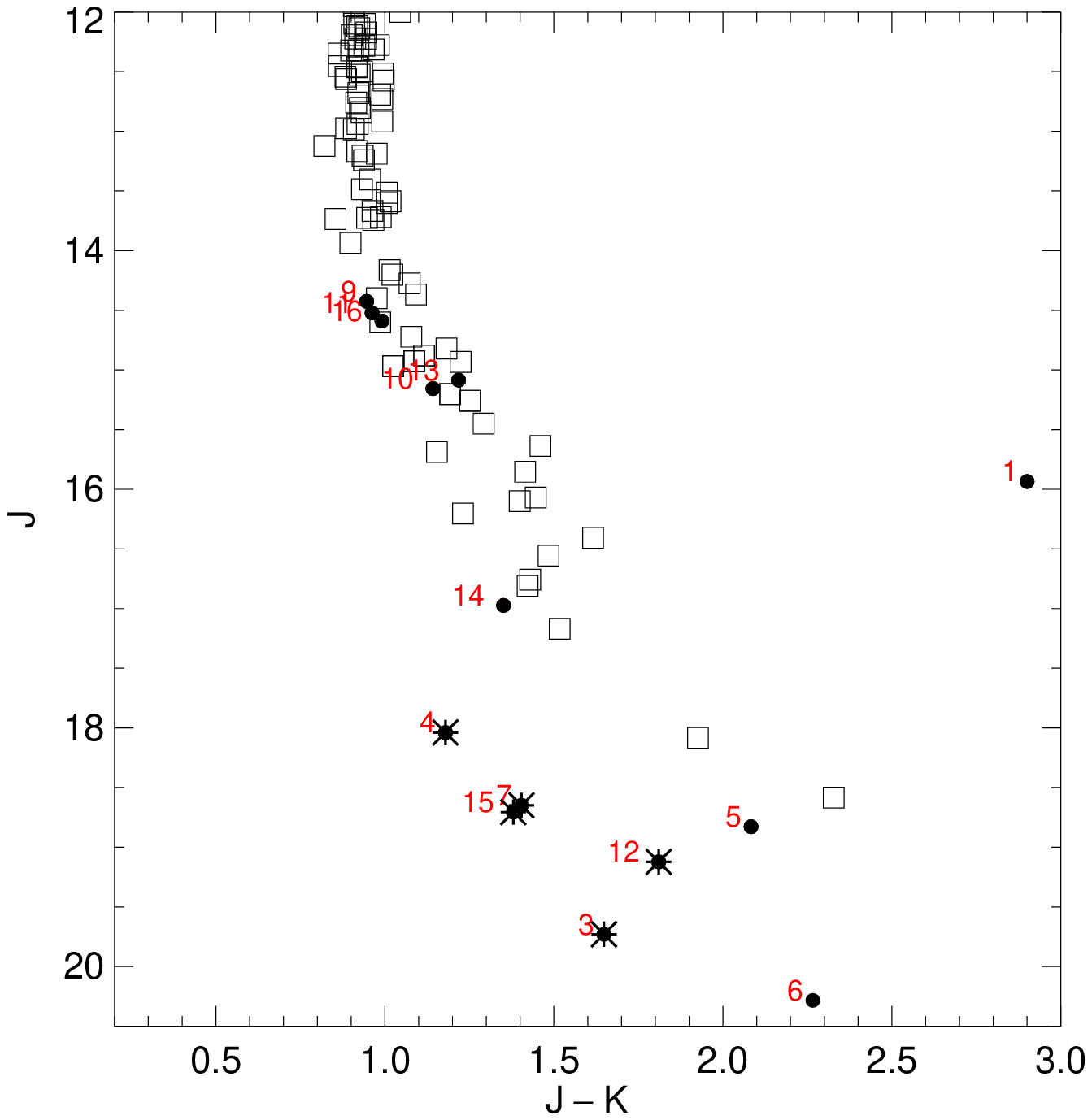}
  \includegraphics[width=0.47\linewidth, angle=0]{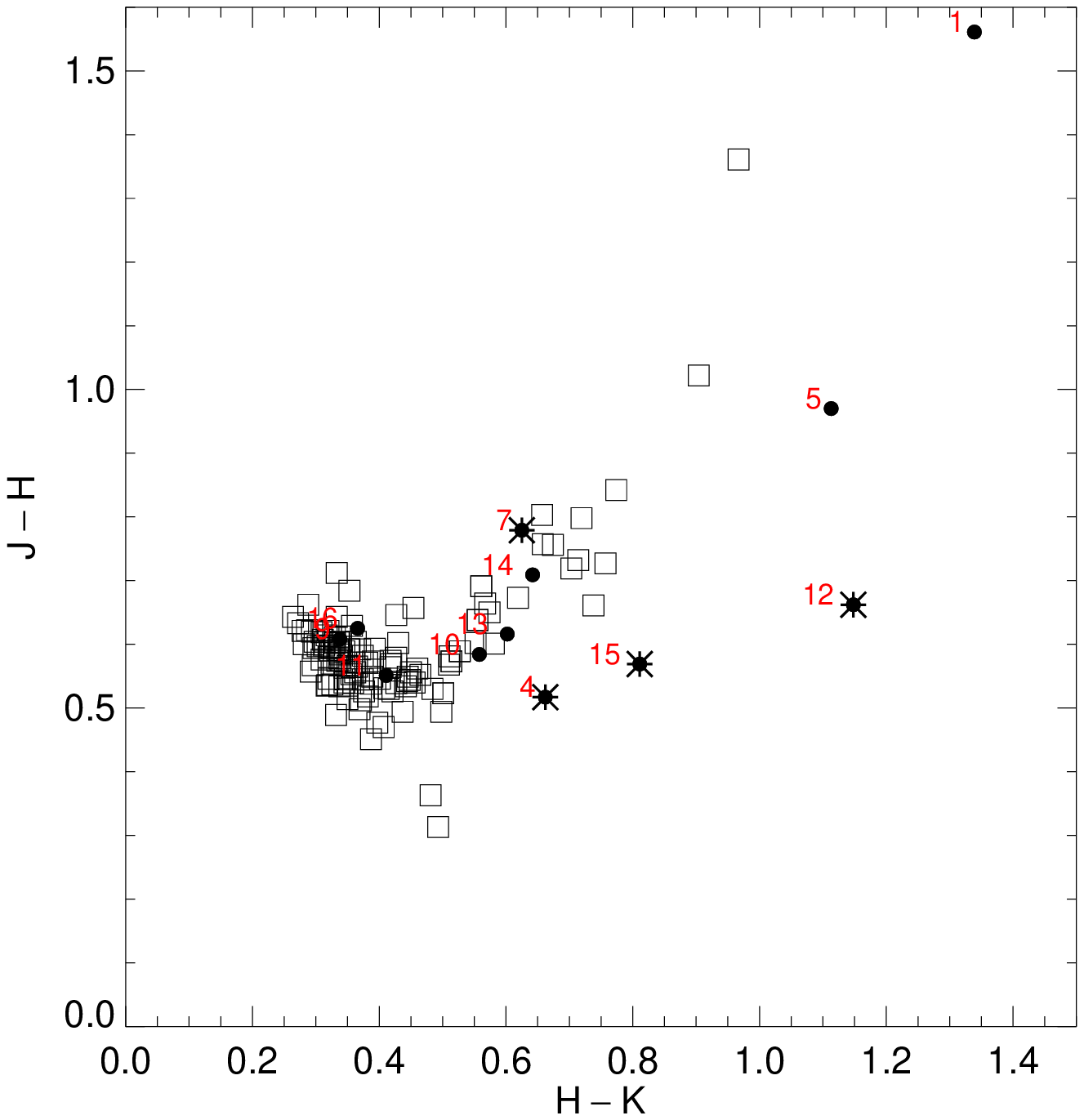}
  \caption{{\it{Top left:}} ($Z-J$,$Z$) colour-magnitude diagram.
{\it{Top right:}} ($Y-J$,$Y$) diagram. 
{\it{Bottom left:}} ($J-K$,$J$) diagram. 
{\it{Bottom right:}} ($H-K$,$J-H$) colour--colour diagram 
The diagrams show the location of the new candidates identified in
our deep $YJ$ survey (filled circles) along with previously known
spectroscopic members of USco from the UKIDSS GCS (open squares).
Photometric non members are marked with black asterisks. The numbers
correspond to the candidates listed in Table \ref{tab_deepUsco:tab_candYJ_new}
in the same order (i.e.\ ordered by right ascension).
}
  \label{fig_deepUsco:cmd_ccd}
\end{figure*}
%

%
%
%
\begin{table*}
 \centering
 \caption[]{Coordinates (in J2000), near-infrared $ZYJ$ magnitudes from the
            deep survey, $Z$ photometry from the Science Verification phase
            (not yet available in GCS DR8),
            $H$ and $K$ from the latest UKIDSS GCS data release, and 
            methane ON and OFF magnitudes with their associated photometric
            error bars for the 16 new candidates. The sources are ordered
            by right ascension and the first column provide an identifier
            reported in the figures and other tables. \\
            $^{a}$ This object remains as photometric candidate \\
            $^{b}$ This object is a field L dwarf spectroscopically \\
            $^{c}$ Classified as quasar spectroscopically.
            }
 \begin{tabular}{@{\hspace{0mm}}r @{\hspace{1mm}}c @{\hspace{2mm}}c @{\hspace{2mm}}c @{\hspace{2mm}}c @{\hspace{2mm}}c @{\hspace{2mm}}c @{\hspace{2mm}}c @{\hspace{2mm}}c @{\hspace{2mm}}c @{\hspace{2mm}}c @{\hspace{2mm}}c @{\hspace{2mm}}c @{\hspace{2mm}}c @{\hspace{2mm}}c c@{\hspace{0mm}}}
 \hline
 \hline
ID & R.A.        &     dec      &  $Z_{\rm deep}$ &  $Y_{\rm deep}$  &  $J_{\rm deep}$  &  $H_{\rm DR8}$   &  $K_{\rm DR8}$      & $Z_{\rm SV}$    & CH$_{\rm 4off}$ & CH$_{\rm 4on}$ \cr
 \hline
 1 & 16:06:30.92 & $-$22:58:15.2 &        ---       & 17.362$\pm$0.006 & 15.935$\pm$0.004 & 14.374$\pm$0.005 & 13.035$\pm$0.003 & 16.658$\pm$0.011 &        ---       \cr
 2 & 16:06:46.45 & $-$22:31:23.8$^{a}$ &        ---       & 21.977$\pm$0.187 & 20.448$\pm$0.081 &        ---       &        ---       &        ---       & 19.562$\pm$0.094 & 19.405$\pm$0.071 \cr
 3 & 16:07:16.98 & $-$22:42:14.1$^{b}$ &        ---       & 21.221$\pm$0.095 & 19.730$\pm$0.045 &        ---       & 18.082$\pm$0.208 &        ---       &        ---       \cr
 4 & 16:07:55.42 & $-$22:33:52.5 & 20.357$\pm$0.027 & 19.220$\pm$0.018 & 18.039$\pm$0.013 & 17.522$\pm$0.070 & 16.860$\pm$0.065 & 20.159$\pm$0.178 &        ---       \cr
 5 & 16:08:12.99 & $-$23:04:31.4 &        ---       & 20.385$\pm$0.053 & 18.828$\pm$0.022 & 17.858$\pm$0.102 & 16.745$\pm$0.065 &        ---       & 17.847$\pm$0.020 & 17.501$\pm$0.013 \cr
 6 & 16:08:35.54 & $-$22:53:11.4 &        ---       & 21.835$\pm$0.171 & 20.284$\pm$0.069 & 18.220$\pm$0.144 & 18.018$\pm$0.196 &        ---       & 19.760$\pm$0.092 & 19.747$\pm$0.061 \cr
 7 & 16:08:42.38 & $-$22:23:25.9 &        ---       & 20.012$\pm$0.030 & 18.649$\pm$0.018 & 17.870$\pm$0.094 & 17.245$\pm$0.091 &        ---       & 17.804$\pm$0.019 & 17.685$\pm$0.008 \cr
 8 & 16:09:00.40 & $-$23:11:51.0$^{c}$ &        ---       & 21.410$\pm$0.112 & 19.751$\pm$0.044 &        ---       &        ---       &        ---       & 19.962$\pm$0.279 & 20.100$\pm$0.204 \cr
 9 & 16:09:03.86 & $-$22:23:22.2 & 15.693$\pm$0.003 & 15.116$\pm$0.002 & 14.425$\pm$0.002 & 13.817$\pm$0.003 & 13.479$\pm$0.003 & 15.633$\pm$0.006 &        ---       \cr
10 & 16:09:05.67 & $-$22:45:16.7 &        ---       & 16.126$\pm$0.003 & 15.156$\pm$0.003 & 14.572$\pm$0.006 & 14.014$\pm$0.005 & 17.341$\pm$0.019 &        ---       \cr
11 & 16:10:00.17 & $-$23:12:19.3 &        ---       & 15.270$\pm$0.002 & 14.522$\pm$0.002 & 13.971$\pm$0.004 & 13.560$\pm$0.004 & 15.947$\pm$0.007 &        ---       \cr
12 & 16:11:05.28 & $-$22:22:57.3 &        ---       & 20.526$\pm$0.051 & 19.123$\pm$0.027 & 18.461$\pm$0.192 & 17.313$\pm$0.109 & 20.000$\pm$ ---  &        ---       \cr
13 & 16:11:27.61 & $-$22:43:33.6 &        ---       & 15.913$\pm$0.003 & 15.085$\pm$0.002 & 14.469$\pm$0.005 & 13.867$\pm$0.005 &        ---       &        ---       \cr
14 & 16:12:44.88 & $-$23:02:14.0 &        ---       & 18.168$\pm$0.009 & 16.973$\pm$0.007 & 16.264$\pm$0.022 & 15.622$\pm$0.025 &        ---       &        ---       \cr
15 & 16:12:57.41 & $-$23:05:55.4 &        ---       & 19.989$\pm$0.032 & 18.706$\pm$0.020 & 18.137$\pm$0.116 & 17.326$\pm$0.119 &        ---       &        ---       \cr
16 & 16:13:30.35 & $-$22:44:06.7 &        ---       & 15.273$\pm$0.002 & 14.590$\pm$0.002 & 13.965$\pm$0.004 & 13.599$\pm$0.004 & 15.840$\pm$0.007 &        ---       \cr
  \hline
 \label{tab_deepUsco:tab_candYJ_new}
 \end{tabular}
\end{table*}

To further constrain the membership of the 16 new candidates, we looked
for additional photometry in $Z$, $H$, and $K$ in the deep 
$Z$-band survey, in the latest release of the UKIDSS GCS (Data Release 8 in 
September 2010) and in the GCS SV database. Two objects are detected in the 
deep $Z$ image and show magnitudes consistent with the photometry extracted 
from the SV database within the error bars 
(Table~\ref{tab_deepUsco:tab_candYJ_new}). Their 
positions in the ($Z-J$,$Z$) diagram \citep[Top, left panel in 
Figure~\ref{fig_deepUsco:cmd_ccd}; see also Figure 3 of][]{lodieu07a} 
place them at the border between the field stars and the cluster sequence. 
Additionally, we extracted $H$ and $K$ photometry from GCS DR8 for 
13 out of 16 new candidates. The other three sources lie in a gap within 
the GCS SV coverage due to control quality rejection in the SV data 
\citep[see Figure 1 in][]{lodieu07a}.
We plotted these 13 sources in the ($Y-K$,$Y$) and ($J-K$,$J$) 
colour--magnitude as well as in the ($H-K$,$J-H$) colour--colour diagram 
(Figure~\ref{fig_deepUsco:cmd_ccd}) along with spectroscopically confirmed 
USco members to assess their membership. Four candidates
(USco J160755.42$-$223352.5; USco J161257.41$-$230555.4;
USco J160842.38$-$222325.9; USco J161105.28$-$222257.3)
are classified as photometric non members because they exhibit blue colours
in several diagrams (asterisks in Figure~\ref{fig_deepUsco:cmd_ccd}). There
is another object (USco J160716.98$-$224214.1) for which only $K$--band data 
is available, suggesting that it is also a photometric non member due to its 
blue infrared colours (asterisk in Figure~\ref{fig_deepUsco:cmd_ccd}). The
remaining eight candidates remain bona-fide members of the USco association.
Finally, we have added a photometric constraint using the $Z$ photometry
from the GCS SV for seven sources out of the 16\@. The brightest candidates 
are confirmed as photometric candidates (top left panel of 
Figure~\ref{fig_deepUsco:cmd_ccd}) while the two faintest sources 
(USco J161105.28$-$222257.3 and USco J160755.42$-$223352.5) are rejected 
due to their blue $Z-J$ colours (note that they were already rejected based 
on their $Y-K$ and $J-K$ colours).

Some of the 16 new candidates are quite bright but were not selected as 
potential members in the SV survey. Seven of them are, however, in the
SV coverage with $YJHK$ photometry and should have been recovered if
indeed members (Table~\ref{tab_deepUsco:tab_candYJ_new}). The two faintest, 
USco J160755.42$-$223352.5 and USco J160842.38$-$222325.9, are photometric 
non members due to their blue infrared colours as discussed above. The 
remaining five are much brighter. One object, USco J161000.17$-$231219.3, 
was included in our AAOmega optical follow--up and classified as a 
spectroscopic non member on the basis of its strong Na{\small{I}} doublet
\citep[Table C.1 in][]{lodieu11a}.
USco J160903.86$-$222322.2 and USco J161330.35$-$224406.7 are clear
proper motion non members based on our estimate from the 2MASS/GCS SV 
crossmatch, a hypothesis corroborated by the US Naval Observatory 
measurements \citep{monet03}. USco J160905.67$-$224517.0 was also classified
as a proper motion non member from our 2007 study but lies very close
to the 2$\sigma$ proper motion selection. Finally, 
USco J160630.92$-$225815.2 appears as a bona-fide photometric and proper 
motion member from our SV study and the new deep survey. The $Y$ and $J$
magnitudes became fainter by approximately 1.4 and 1.0 mag, respectively,
between the SV and deep survey epochs. This source was also included in
our AAOmega multi--fibre spectroscopic follow--up \citep{lodieu11a} and 
its spectrum suggests that it is a young active star with strong H$\alpha$, 
Helium, N{\small{II}}, and S{\small{II}} emission lines and whose continuum 
is rising steeply from 6000 to 9000\,\AA{}. Overall we have rejected
sources which do not satisfy at least two colours criteria or a colour
and proper motion or spectroscopic criteria. The final membership is given 
in the last column of Table~\ref{tab_deepUSco:PM_values} and a note has
been added to Table \ref{tab_deepUsco:tab_candYJ_new} for the three sources 
not in Table~\ref{tab_deepUSco:PM_values}.
The other bright candidates identified in this paper could not have been
recovered by our SV study because we imposed a simultaneous detection in
all four $YJHK$ passbands \citep{lodieu07a}.

The next step was to crossmatch this catalogue of new candidates with 
the methane imaging data from WIRCam to look at the cluster sequence
with respect to the methane colour (Figure~\ref{fig_deepUsco:cmd_CH4J}).
We crossmatched the full deep USco survey and the methane survey
with a matching radius of 2~arcsec. The total number of common sources
is 65,152 over one square degree. We also crossmatched the known 
spectroscopic members with the full methane survey, returning only one 
source, USco J160918.69$-$222923.7 \citep[L1;][]{lodieu08a} out of the 
30 members covered by WIRCam (green open square in 
Figure~\ref{fig_deepUsco:cmd_CH4J}). We identified another object with 
methane photometry, USco J160956.34$-$222245.6, classified as a field L2 
dwarf by \citet{lodieu08a}. Its methane colour is bluer than the young 
L1 type belonging to USco. We should mention that the CH$_{\rm 4off}$ and 
CH$_{\rm 4on}$ images 
saturate at about 16 mag, corresponding to roughly 17.7 and 16.7 mag in $Y$ 
and $J$, respectively. As a consequence, none of the bright members and new 
candidates will be retrieved in the methane survey. We found that five 
of the seven faintest ($Y$\,$\geq$\,20 mag) new candidates from the $YJ$ 
survey have methane photometry (filled red dots in Figure~\ref{fig_deepUsco:cmd_CH4J}.

%
%
\section{Proper motions}
\label{deepUsco:select_PM}

In this section, we take advantage of the large epoch difference 
between the GCS SV and the deep $YJ$ survey to estimate proper motions for 
cluster member candidates listed in Table~\ref{tab_deepUsco:tab_candYJ_new}. 
The baseline is of the order of three years: the GCS SV data were
obtained on 12 and 19 April 2005 whereas the deep survey was carried out
on 4 and 5 May 2008 taking the $J$-band observations as reference.
The baseline between the deep $YJ$ survey and methane observations is 
less than two months, too small to infer any reliable membership based 
on proper motion for the T dwarf candidates in 
Table~\ref{tab_deepUSco:CH4_candidates}.
The typical accuracy on the proper motion measurement is better than
10 mas/yr for the sources with photometric detections of 20$\sigma$ or more. 
This centroiding accuracy decrease linearly with the signal-to-noise, implying 
that sources with 10$\sigma$ and 5$\sigma$ detections have proper motion 
errors of the order of 14--18 mas/yr (depending on the epoch difference; 3 or
4 years) and 30 mas/yr for objects fainter than $J$ = 19 mag using the 4 year
baseline, respectively. Thus, the faintest candidate identified in this
survey, USco J160835.54$-$225311.4, and classified as a proper motion non 
member is consistent with the USco mean motion within the error bars
(Table \ref{tab_deepUSco:PM_values}).

There are 13 $YJ$ candidates with GCS SV astrometry. We computed
their proper motions using the GCS SV as first epoch and the deep $YJ$
survey as second epoch. The proper motions in right ascension and
declination are listed in Table~\ref{tab_deepUSco:PM_values}.
We show that 11 out of 13 photometric candidates exhibit proper motions
within 3$\sigma$ of the USco mean proper motion ($-$11,$-$25 mas/yr),
implying that they remain as probable members.
We should emphasise that the same conclusion is reached when using
the GCS DR8 instead of the deep $YJ$ as a second epoch.
One of the proper motion non members, USco J161257.41$-$230555.4,
was classified as a photometric non member earlier in our analysis.

The one year baseline between the deep $YJ$ survey and the GCS DR8 suggests
that USco J160716.98$-$224214.1 may be a proper motion non member despite
the large uncertainty. Adding this argument to its rejection as
a photometric member (Section~\ref{deepUsco:select_Ldwarfs}), this object
very likely does not belong to the USco association.

%
%
%
%
\begin{table*}
 \centering
 \caption[]{Coordinates, observing dates, and estimated proper motions
            in right ascension and declination for the 13 candidates
            common to the deep $YJ$ survey, the GCS DR8, and the GCS
            SV databases. The last column provide additional comments 
            based on additional data discussed in the text: the terminology
            is as follows: NM stands for non member; PM for proper motion;
            photNM for photometric non member; and L11a for \citet{lodieu11a}.}
 \begin{tabular}{@{\hspace{0mm}}r c c c c c c c c@{\hspace{0mm}}}
 \hline
 \hline
ID & R.A. (deep) &  dec (deep)   &  Epoch deep  &  Epoch DR8  & Epoch SV & $\mu_{\alpha}cos{\delta}$ & $\mu_{\delta}$ & Comments \cr
 \hline
 9 & 16:09:03.86 & $-$22:23:22.2 & 2008.34153 & 2009.33973 & 2005.29863 &   $-$23.57 &   $-$39.19 & PM\_NM  \cr
11 & 16:10:00.17 & $-$23:12:19.3 & 2008.34426 & 2008.40713 & 2005.29863 &    $-$6.27 &    $+$2.42 & NM\_L11a \cr
16 & 16:13:30.35 & $-$22:44:06.7 & 2008.34426 & 2009.33973 & 2005.27945 &    $-$7.91 &    $-$9.57 & PM\_NM  \cr
13 & 16:11:27.61 & $-$22:43:33.6 & 2008.34426 & 2009.33973 & 2005.27945 &    $-$4.27 &    $-$8.92 & \cr
10 & 16:09:05.67 & $-$22:45:16.7 & 2008.34153 & 2009.33973 & 2005.29863 &    $-$0.88 &    $-$4.58 & PM\_borderline \cr
 1 & 16:06:30.92 & $-$22:58:15.2 & 2008.34153 & 2008.40713 & 2005.29863 &   $+$2.31  &    $-$1.81 &    \cr
14 & 16:12:44.88 & $-$23:02:14.0 & 2008.34426 & 2008.40713 & 2005.27945 &   $-$13.11 &    $-$4.20 & \cr
 4 & 16:07:55.42 & $-$22:33:52.5 & 2008.34153 & 2009.33973 & 2005.29863 &   $-$24.46 &    $-$6.83 & photNM \cr
15 & 16:12:57.41 & $-$23:05:55.4 & 2008.34153 & 2008.40713 & 2005.27670 &   $+$92.03 &    $+$8.75 & photNM \cr
 7 & 16:08:42.38 & $-$22:23:25.9 & 2008.34153 & 2009.33973 & 2005.29863 &   $-$11.71 &   $-$24.37 & photNM \cr
 5 & 16:08:12.99 & $-$23:04:31.4 & 2008.34153 & 2008.40713 & 2005.29863 &   $-$10.50 &   $-$20.09 & \cr
12 & 16:11:05.28 & $-$22:22:57.3 & 2008.34426 & 2009.33973 & 2005.27945 &    $-$3.69 &   $-$14.03 & photNM \cr
 6 & 16:08:35.54 & $-$22:53:11.4 & 2008.34153 & 2009.33973 & 2005.29863 &   $+$44.36 &    $-$3.29 & PM\_NM \cr
  \hline
 \label{tab_deepUSco:PM_values}
 \end{tabular}
\end{table*}
%
%
%
\section{Spectroscopic follow-up}
\label{deepUsco:XSHOOTERspec}

In this section, we analyse the optical and near--infrared cross--dispersed
spectra of four candidates identified in the 0.95 square degree
area in the central part of the USco association common to the $YJ$
and methane surveys.

%
%
%
\begin{figure*}
  \centering
  \includegraphics[width=\linewidth, angle=0]{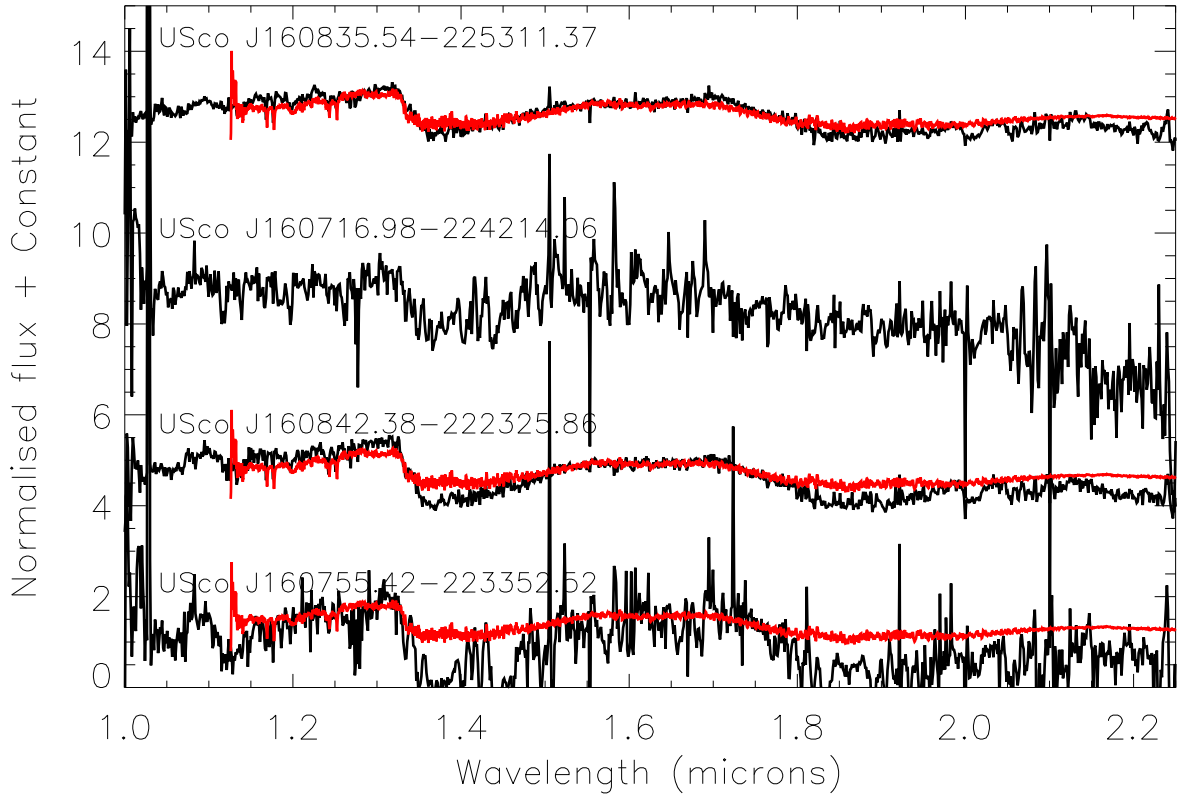}
  \caption{X-shooter spectra of four photometric candidates covering the 
1.0--2.5 micron wavelength. The spectra have been normalised at 1.265 microns 
and smoothed by a factor of 20 with using a weighted sum procedure, resulting 
in an imrpoved signal-to-noise and a resolution of $\sim$250 in the 
near--infrared. From bottom to top,
we show USco J160716.98$-$224214.06, USco J160755.42$-$223352.52,
USco J160835.54$-$225311.37, and USco J160842.38$-$222325.86\@.
Overplotted in red as a comparison is the near--infrared spectrum of the 
known field L5 dwarf, 2MASS J15074769$-$1627386\@.
   }
  \label{fig_deepUsco:final_NIR_spectra}
\end{figure*}
\subsection{Spectral analysis}
\label{deepUsco:XSHOOTERspec_Analysis}

In this section, we discuss the spectral properties of the four photometric
candidates whose spectral energy distribution resembles that of L-type dwarfs.

The X--Shooter spectra of USco J160755.42$-$223352.52 and 
USco J160842.38$-$222325.86 have much better signal--to--noise than
the other two sources. In the visible arm, the spectrum
of USco J160842.38$-$222325.86 appears redder towards longer wavelengths,
suggesting a later spectral type. The same trend is seen in the $J$--band
region of the near--infrared arm whereas the $H$--band region appears quite 
similar. The near--infrared spectra of the two fainter photometric candidates 
are very similar to the spectrum of USco J160755.42$-$223352.52 despite the
low signal--to--noise and the presence of many sky lines not well removed
by the current version of the X-shooter pipeline. A more quantitative
comparison is unfortunately not possible at this stage.

All of the four photometric candidates discussed here look like field
mid- to late-L dwarfs. For example, they do not harbour the peaked-shape 
in the $H$--band typical of young L--type brown dwarfs
identified in star forming regions \citep[e.g.][]{allers07,lodieu08a},
pointing towards age later than a few hundred Myrs. We compared our spectra 
to low resolution near--infrared of ``template'' brown dwarfs observed with 
the NIRSPEC spectrograph \citep{mclean03}. We found that the $J$ and $H$ 
regions of our near--infrared spectra 
are best fit by the NIRSPEC spectra of 2MASS J15074769$-$1627386
\citep[L5;][]{reid00,kirkpatrick00,knapp04} overplotted in red on the
best signal-to-noise spectra in Fig.\ \ref{fig_deepUsco:final_NIR_spectra}. 
The uncertainty on the spectral type is of the order of one subclass, 
suggesting that the four object are L4--L6 field dwarf. The optical spectrum 
of USco J160755.42$-$223352.52  confirm such a spectral type whereas
USco J160842.38$-$222325.86 may be slightly later by 1 or 2 subclass, 
possibly L8 according to its optical shape.

Moreover, we measured a few gravity--sensitive doublets, including
the K{\small{I}} doublets at 1.168/1.177 and 1.243/1.254 microns and
the Na{\small{I}} doublet at 1.268 microns, to add 
constraints on the age of these L-type candidates and their membership of 
the USco association. These doublets have been extensively used in the 
literature to disentangle field brown dwarfs from young substellar objects 
\citep[e.g.][]{mcgovern04,allers07,cushing05,lodieu08a,lafreniere10a,bonnefoy10a}.
We resolved the K{\small{I}} doublets at the resolution of the X--Shooter 
spectra but not the Na{\small{I}} doublet which is also affected by a
strong sky line which removes about 10\% of the flux in this region. 
We measured their pseudo--equivalent widths after smoothing the $J$--band 
spectra by a factor of 20 with a weighted sum procedure\@. We used the task 
{\tt{SPLOT}} to carry out such measurements. Typical error bars on the 
measurements are 1--2\,\AA{} for the two brightest sources and 
$\sim$2--3\,\AA{} for the faintest candidates. 
The values reported in 
Table~\ref{tab_deepUSco:NIR_spectral_indices} for these three doublets 
confirm that these sources are old L dwarfs because of the presence of 
strong equivalent widths. Hence, we classify these four photometric 
candidates as mid- to late-L field dwarfs and reject them as members of 
the USco association.

%
%
%
\begin{table}
 \centering
 \caption[]{Pseudo-equivalent widths of gravity-sensitive doublets
           for four photometric candidates identified in USco.
          We give the equivalent widths in \AA{} for each component of the
          three doublets: K{\small{I}} at 1.168/1.177 and 1.243/1.254 microns
          and the sum of the Na{\small{I}} doublet at 1.268 microns.}
 \begin{tabular}{@{\hspace{0mm}}c c c c@{\hspace{0mm}}}
 \hline
 \hline
USco J\ldots{}    &  K{\small{I}} & K{\small{I}} & Na{\small{I}}  \\
                  &  1.168/1.177  & 1.243/1.254  &      1.268     \\
 \hline
160716.98$-$224214.06 & 12.6$+$13.4 & 12.6$+$15.0 & 13.7 \\ 
160755.42$-$223352.52 &  9.2$+$8.2  &  8.4$+$8.4  &  6.5 \\ 
160835.54$-$225311.37 & 19.4$+$9.3  &  7.3$+$7.5  & 12.7 \\ 
160842.38$-$222325.86 &  9.2$+$12.2 &  7.4$+$6.2  &  7.1 \\ 
  \hline
 \label{tab_deepUSco:NIR_spectral_indices}
 \end{tabular}
\end{table}
%

%
%
%
\begin{figure}
  \centering
  \includegraphics[width=\linewidth, angle=0]{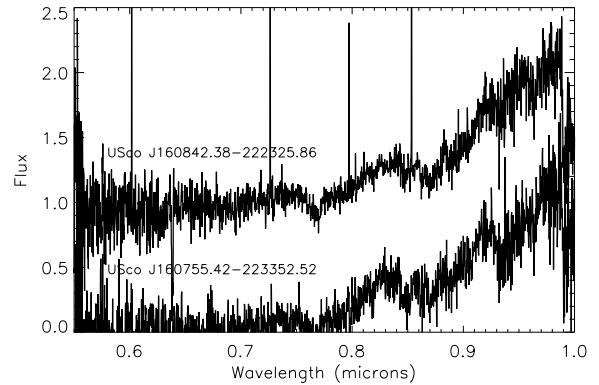}
  \caption{X-shooter spectra of two photometric candidates covering
the 0.6--1.0 micron wavelength. The spectra have been normalised at
9200\,\AA{} and smoothed by a factor of 20 with a weighted sum procedure, 
resulting to a resolution of $\sim$200 in the optical.}
  \label{fig_deepUsco:final_VIS_spectra}
\end{figure}
\subsection{Properties of the quasar}
\label{deepUsco:XSHOOTERspec_QSO}

\begin{figure}
  \centering
  \includegraphics[width=\linewidth, angle=0]{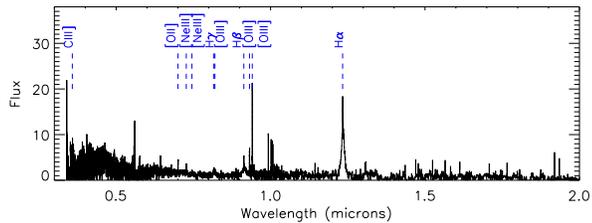}
  \caption{X-shooter optical and near-infrared spectrum of the quasar at 
redshift $\sim$0.88 found during our photometric search for low-mass brown 
dwarf members in USco. The main emission lines are marked.
   }
  \label{fig_deepUsco:QSOspectrum}
\end{figure}

In this section we discuss the properties of the quasar found in our search
for ultracool brown dwarfs in USco. It is interesting to see that a methane
search for substellar objects can be contaminated by quasars at low redshift
\citep{goldman10}.
Its spectrum is shown in Fig.\ \ref{fig_deepUsco:QSOspectrum}

In the extracted spectrum of USco J160900.40--231150.95, we identify strong,
broad Balmer emission lines typical of quasars. As shown in  
Fig.\ \ref{fig_deepUsco:QSOspectrum}, narrower Oxygen, Neon, and Carbon lines
are found in addition to the Balmer lines, and from these lines we determine
an average redshift of 0.8789$\pm$0.0002. 

Compared to a composite spectrum of quasars from the SDSS \cite{VandenBerk01}
the H$\beta$ and [OIII] $\lambda\lambda$4959,5007 lines have similar 
strengths relative to the continuum flux, while the H$\alpha$ emission is 
stronger by a factor of $\sim$8. Apart from the strong lines, the continuum 
slope is consistent with the average quasar spectral slope. These 
characteristics arise when the emission from the quasar broad line region 
is partly absorbed by dust in a non-uniform dust configuration, which allows 
the quasar continuum emission to pass though relatively un absorbed. 
Since H$\alpha$ falls within the $J$ band at the redshift of the quasar, the 
red $Y-J$ colour is dominated by the excess emission from this line.

%
%
%
%

%
%
\section{Discussion: testing the fragmentation limit at 5 Myr}
\label{deepUsco:discussion}

In this section, we summarise the results and put our results into context 
discussing the possible turn down of the mass function in the context of 
theory of fragmentation limit.

Out of the 16 candidates identified in the deep $YJ$ survey presented here, 
we are left with four candidates with $YJHK$ photometry and proper motions
consistent with the USco mean motion. Three of these four candidates
are brighter than the GCS limit and outside the GCS coverage; one being 
fainter and new. We identified another photometric candidates based on 
its $Y-J$ colour only as well as five potential young T dwarfs in the methane 
survey without optical counterpart.

Our new deep $YJ$ survey is approximately 2 magnitudes deeper in both $Y$ and 
$J$ filters than the 5$\sigma$ completeness limits of the UKIDSS GCS\@.
We identified one candidate with a magnitude comparable to the depth of the
GCS and another at the limit of the deep $YJ$
survey, suggesting that this new study has revealed at most one new low-mass
brown dwarf in USco. Indeed we do not confirm any new candidates in the
$Y$ = 20--22 mag range and identified only one new photometric candidate
(not counting the T dwarf candidates that require additional follow-up).
The lowest mass brown dwarf identified in the GCS and confirmed 
spectroscopically \citep{lodieu08a} has an effective temperature of 
$\sim$1800 K and a mass of $\sim$8 M$_{\rm Jup}$ according to the DUSTY models
\citep{chabrier00c}. If we consider that our new survey is roughly two 
magnitudes deeper, this would correspond to masses of 3--4 M$_{\rm Jup}$ 
following predictions from the dust-free COND models \citep{baraffe02}.
Overall, this result is puzzling because we find many candidates that look 
like USco photometric members in various colour-magnitude diagrams which 
we rejected later on the basis of their astrometry and/or spectroscopy.

In our studies of the central region of the USco association, we
confirmed spectroscopically 14 members with masses in the 0.02--0.01 M$_{\odot}$
over 6.5 square degrees \citep{lodieu07a,lodieu08a,lodieu11a}. This mass range 
corresponds roughly to $J$ = 15--17.2 mag according to the DUSTY models 
of the Lyon group \citep{chabrier00c}. We want to address the following 
question: how many new brown dwarfs do we expect down to the limit of our 
deep $YJ$ survey covering 1.7 square degree. If we assume a constant mass
function in the next mass bin chosen as 0.01--0.005 M$_{\odot}$ (corresponding
to $J$ $\sim$ 17.2--21.5 mag), we would expect 14/6.5*1.7=3.7 members. However,
if we extrapolate the field mass function in the same mass bin considering
that we found a factor of three difference in the number of objects compared
to our spectroscopic mass function \citep{lodieu11a}, we would expect 1.2
objects in our deep survey. Hence, the one photometric candidate we extracted
in our deep survey is consistent with the low limit of those predictions.

We estimated the number of field T0--T8 dwarfs expected in our deep $YJ$ survey 
covering 1.7 square degrees down to $J$ = 21.5 mag. \citet{pinfield08} found 
26--44 T0--T8 dwarfs in 280 square degrees surveyed by the UKIDSS Large Area 
Survey down to $J$ = 19 mag. Scaling these numbers to the area covered by our
$YJ$ survey and the depth, we estimate that our survey may be contaminated by 
5.0--6.6 to 8.4--11.1 field T dwarfs. However, if we consider only the 
methane coverage which is designed to focus on T dwarfs and which is shallower 
than the deep $YJ$ survey ($H$ $\sim$ 20.2--20.3 mag), we would expect 
1--2 T dwarfs compared to our five candidates listed in 
Table \ref{tab_deepUSco:CH4_candidates}. These numbers suggest a high level 
of contamination of at least 20 to 40\% among our T dwarf candidates.
This result is in line with the classification of four of the faint photometric 
candidates as spectroscopic non members.

Although we are missing additional membership constraints on the potential 
T dwarf candidates presented here, we are in a position to discuss a
possible turn down in the mass function as we approach the fragmentation limit 
in USco as our deep survey has not revealed any new spectroscopic late-L or 
T dwarf. Independent works in $\sigma$ Orionis by \citet{bihain09} and 
\citet{penya11a} presented deep infrared and methane surveys with similar 
depths in terms of mass as our study where the numbers of T-type member 
candidates seem to fall down below $\sim$6 Jupiter masses because none of the 
T dwarfs announced to date in this region has been unambiguously confirmed 
astrometrically and spectroscopically. However, the spectral sequence in 
$\sigma$\,Orionis extends down to late-L dwarfs \citep{barrado01c,martin01a} 
while our USco sequence currently stops at early-L dwarfs or possibly mid-L 
if we include the planetary-mass companion found by \citet{lafreniere10a} 
around a solar analog despite the possible excess of brown dwarfs seen in USco 
\citep{preibisch02,slesnick08,lodieu11a}. Our survey seems to favour a
turn down in the USco mass function below 10 Jupiter masses but a wider 
survey as deep the one presented here is needed to improve the statistics
and confirm this possible turn down. We cannot yet argue that we have reached
the fragmentation limit in USco. To conclude, T-type members
with masses below a few Jupiter masses seem to be rare at very young ages 
($<$10 Myr), placing an upper limit on the smallest fragments that star 
formation processes can form \citep{low76,rees76,boss01}.

%
%
\section{Summary and outlook}
\label{deepUsco:summary}

We have presented the outcome of deep photometric survey of 0.95 square
degree in the USco association combining near-infrared and methane imaging.
We identified 16 new candidates in 1.7 square degrees surveyed in
$Y$ and $J$, leaving four probable members including at the limit of our
survey after applying additional photometric and astrometric constraints 
from various surveys taken at different epochs. Near-infrared spectroscopy
of four of the faintest candidates has shown that their are old field
mid-L dwarfs contaminating our photometric search while the other is a
quasar at low redshift. Moreover, we also extracted five potential methane 
T dwarf candidates without optical counterparts and reasonable $Y-J$ colours. 
The lack of new spectroscopic members in USco in a survey two magnitudes 
deeper than the UKIDSS GCS points towards a turn down in the USco mass
function or may suggest that we have reached the fragmentation limit in the 
association.

Spectroscopy of these faint T dwarf candidates is difficult
with ground-based facility, suggesting that these sources may be ideal
targets for the James Webb Space Telescope and the future E-ELT\@.
Proper motion confirmation is a more affordable option to confirm
the membership of these T dwarfs to the USco association very soon
with a second epoch. Furthermore, the advent of the Visible Infrared
Survey Telescope for Astronomy \citep[VISTA;][]{emerson01} will allow to
cover larger areas in smaller numbers of pointing to improve the statistics 
and confirm (or otherwise) that we have reached the fragmentation limit 
in USco.

%
%
\section*{Acknowledgments}
NL was funded by the Ram\'on y Cajal fellowship number 08-303-01-02
and the national program AYA2010-19136 funded by the Spanish ministry of
science and innovation. 
NL thanks Lo\"ic Albert and Patrick Hudelot for their help in the 
preparation of the observing blocks and the data reduction, respectively.
LS thanks NL for funding her one week stay at the IAC, Tenerife.

This work made use of data taken with various telescopes and instruments
in visitor and service mode. The $Y,J$ infrared data with Wide Field Camera
(WFCAM) on the United Kingdom Infrared Telescope (UKIRT), operated by the 
Joint Astronomy Centre (JAC) on behalf of the Science and Technology 
Facilities Council (SFTC) of the United Kingdom (UK).
The methane ON and OFF data are based on observations obtained with WIRCam, 
a joint project of CFHT, Taiwan, Korea, Canada, France, and the 
Canada France Hawaii Telescope (CFHT) which is operated by the National 
Research Council (NRC) of Canada, the Institute National des Sciences de 
l'Univers of the Centre National de la Recherche Scientifique of France, 
and the University of Hawaii.
We acknowledge the Terapix team at the Institut d'Astrophysique de Paris 
that produced the images and catalogues used in this work, with special 
thanks to Patrick Hudelot.
The X-Shooter spectroscopy is based on observations made with ESO telescopes 
at the La Silla Paranal Observatory under programme ID 385.C-0950 in service 
mode.

This research has made use of the Simbad and Vizier databases, operated
at the Centre de Donn\'ees Astronomiques de Strasbourg (CDS), and
of NASA's Astrophysics Data System Bibliographic Services (ADS).

This publication makes use of data products from the Two Micron
All Sky Survey (2MASS), which is a joint project of the University
of Massachusetts and the Infrared Processing and Analysis
Center/California Institute of Technology, funded by the National
Aeronautics and Space Administration and the National Science Foundation.

This publication makes use of data products from the Wide-field Infrared 
Survey Explorer, which is a joint project of the University of California, 
Los Angeles, and the Jet Propulsion Laboratory/California Institute of 
Technology, funded by the National Aeronautics and Space Administration.

%
%
\bibliographystyle{mn2e}
\bibliography{../../AA/mnemonic,../../AA/biblio_old}


\bsp

\label{lastpage}

\end{document}